%% file: main.tex
\documentclass[
  reprint,
  amsmath,amssymb,
  aps,
  prb,     
  floatfix
]{revtex4-1}

\usepackage{graphicx}
\usepackage{dcolumn}
\usepackage{bm}
\usepackage{physics}
\usepackage{amsmath}
\usepackage{amsthm}
\usepackage{xcolor}
\usepackage{wrapfig}
\usepackage{ulem}
\usepackage{booktabs}
\usepackage{gensymb}
\usepackage{hyperref}
\usepackage{orcidlink}
\usepackage{cleveref}
\usepackage{nicefrac}
\usepackage{fancyhdr}
\usepackage{tabularx}
\usepackage{xr}
\usepackage{float}
\usepackage{enumitem}

\usepackage[caption=false,position=top]{subfig}
\usepackage{multirow}
\usepackage{textcomp}

\begin{document}

\preprint{APS/123-QED}

\title{Improving the lifetime of aluminum-based superconducting qubits through atomic layer etching and deposition}

\author{Neha Mahuli\,\orcidlink{0000-0001-8218-5536}}
\altaffiliation{Corresponding email address: mahuli@amazon.com}
\affiliation{AWS Center for Quantum Computing, Pasadena, CA 91106, USA}
\author{Joaquin Minguzzi}
\affiliation{AWS Center for Quantum Computing, Pasadena, CA 91106, USA}
\author{Jiansong Gao}
\affiliation{AWS Center for Quantum Computing, Pasadena, CA 91106, USA}
\author{Rachel Resnick}
\altaffiliation{Current affiliation: Google Research}
\affiliation{AWS Center for Quantum Computing, Pasadena, CA 91106, USA}
\author{Sandra Diez}
\altaffiliation{Current affiliation: National Institute of Standards and Technology}
\affiliation{AWS Center for Quantum Computing, Pasadena, CA 91106, USA}
\author{R Cosmic}
\affiliation{AWS Center for Quantum Computing, Pasadena, CA 91106, USA}
\author{Guillaume Marcaud\,\orcidlink{0000-0002-4694-7262}}
\affiliation{AWS Center for Quantum Computing, Pasadena, CA 91106, USA}
\author{Matthew Hunt}
\affiliation{AWS Center for Quantum Computing, Pasadena, CA 91106, USA}
\author{Loren Swenson}
\affiliation{AWS Center for Quantum Computing, Pasadena, CA 91106, USA}
\author{Jefferson Rose}
\affiliation{AWS Center for Quantum Computing, Pasadena, CA 91106, USA}
\author{Oskar Painter\,\orcidlink{0000-0002-1581-9209}}
\affiliation{AWS Center for Quantum Computing, Pasadena, CA 91106, USA}
\author{Ignace Jarrige\,\orcidlink{0009-0006-7134-7812}}
\altaffiliation{Corresponding email address: jarrige@amazon.com}
\affiliation{AWS Center for Quantum Computing, Pasadena, CA 91106, USA}

\date{\today}

\begin{abstract}

We present a dry surface treatment combining atomic layer etching and deposition (ALE and ALD) to mitigate dielectric loss in fully fabricated superconducting quantum devices formed from aluminum thin films on silicon. The treatment, performed as a final processing step prior to device packaging, starts by conformally removing the native metal oxide and fabrication residues from the exposed surfaces through ALE before \textit{in situ} encapsulating the metal surfaces with a thin dielectric layer using ALD. We measure a two-fold reduction in loss attributed to two-level system (TLS) absorption in treated aluminum-based resonators and planar transmon qubits. Treated transmons with compact capacitor plates and gaps achieve median $Q$ and $T_1$ values of $3.69 \pm 0.42 \times 10^6$ and $196 \pm 22$~$\mu$s, respectively. These improvements were found to be sustained over several months. We discuss how the combination of ALE and ALD reverses fabrication-induced surface damages to significantly and durably improve device performance via a reduction of the TLS defect density in the capacitive elements.

\end{abstract}

\maketitle

\section{Introduction}
Superconducting transmons are a leading platform for building a scalable, fault-tolerant quantum computer capable of solving problems currently intractable with classical computing architecture \cite{wendin2017quantum}. In the early stages of development, transmon performance improved primarily through design optimizations, such as  reducing dielectric participation in critical surfaces \cite{wenner2011surface, wang2015surface, gambetta2016investigating} and enhancing readout via Purcell filtering \cite{reed2010fast, jeffrey2014fast}. More recently, the focus has shifted toward enhancing the quality of materials, a strategy that has already shown great promise in mitigating two-level system (TLS) losses and improving transmon energy-relaxation times \cite{murray2021material, place2021new, bland20252d}.

Among the main dielectric regions in superconducting circuits, the metal-air (MA) and substrate-air (SA) interfaces are particularly susceptible to degradation during the fabrication process. These interfaces are often characterized by polymeric residues, interstitial contaminants, morphological damage, and thickened oxide layers, resulting in a larger loss tangent compared to other interfaces \cite{woods2019determining}. This stresses the need to develop post-fabrication surface treatment methods that alleviate or eliminate these effects without introducing new damage. 

Recent efforts to reduce dielectric loss at the MA interface have centered on three primary approaches: (i) exploring new ground-layer materials that form low-loss native oxides \cite{place2021new}, (ii) chemically etching defect-rich surface layers \cite{altoe2022localization, crowley2023disentangling, kopas2024enhanced, lozano2024low}, and (iii) capping the top surface of the ground layer with a lower-loss superconductor, or a noble metal to prevent oxidation \cite{zhou2024ultrathin, bal2024systematic, karuppannan2024improved, chang2025eliminating}. In (iii), surface capping was performed on the blanket ground layer prior to fabrication, as subsequent etching exposed the sidewalls of the superconductor. As a result, amorphous oxides and process-induced defects on the sidewalls likely dominated the dielectric loss in these capped devices \cite{chang2025eliminating}. Here, we present an alternative surface engineering approach that overcomes this limitation by treating all exposed surfaces of fully fabricated devices using a combination of atomic layer etching and deposition (ALE and ALD) as a final surface modification step prior to device packaging. 

Both ALE and ALD rely on a self-limiting, sequential gas-phase surface reaction mechanism to respectively remove and form material layers with monolayer precision and a high conformality compatible with three-dimensional structures. By applying these techniques to aluminum (Al)-based coplanar waveguide (CPW) resonators and planar transmons, we show that fabrication-related contamination of the Al and silicon (Si) substrate surfaces is substantially reduced and the lossy native Al oxide layer is selectively replaced with a thin and stoichiometric Al$_{2}$O$_{3}$ capping layer. These modifications to the MA and SA interfaces correlate with a 50\% reduction in dielectric loss in the resonators and a two-fold increase in both the transmon median quality factor ($Q$) and energy relaxation times ($T_{1}$), reaching median $Q$ and $T_1$ values of $3.69 \pm 0.42 \times 10^6$ and $196 \pm 22$~$\mu$s, respectively. These improvements are achieved for several fabricated chips, and across a frequency range spanning $2.2$-$4.4$~GHz. This study shows that the \textit{in situ} combination of ALE and ALD serves as an effective and scalable surface engineering strategy for Al-on-Si superconducting quantum circuits, with potential applications across a myriad of architectures and materials.

\section{Results and Discussion}

\subsection{Surface Engineering Effects on Resonator Loss Tangent}

\begin{figure}[htbp]
    \centering
    \includegraphics[width=0.48\textwidth]{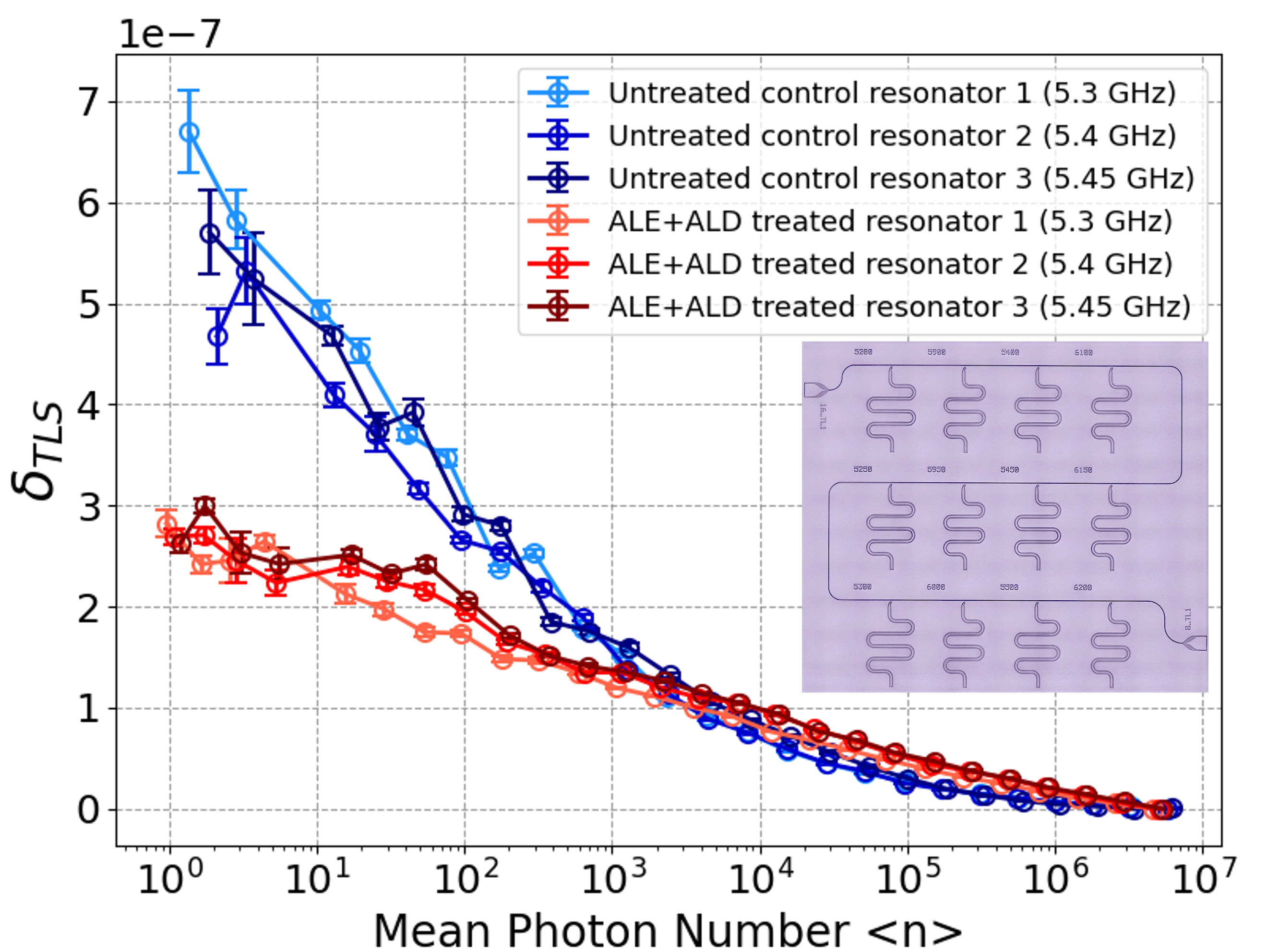}
    \caption{$\delta_{\text{TLS}}$ as a function of power for three representative resonators, comparing untreated and ALE+ALD treated chips. Inset: Micrograph of a CPW resonator chip.}
    \label{fig:Resonator}
\end{figure}

Superconducting microwave resonators provide an effective platform for evaluating materials-induced dielectric loss, minimizing the design, fabrication, and measurement overhead compared to transmons \cite{mcrae2020materials, marcaud2025low}. $\lambda$/4-CPW resonators were fabricated using $100$-nm thick Al films on high-resistivity ($>20\,\mathrm{k\Omega\cdot cm}$) silicon substrates, with trace and gap widths of 30 $\mu$m each. Each chip contains 12 resonators with frequencies ranging from 5.2 to 6.2 GHz, inductively coupled to the transmission line, as shown in the inset of Figure \ref{fig:Resonator}. We performed $S_{21}$ transmission measurements in a dilution refrigerator at a base temperature of $10$~mK and the intrinsic quality factors ($Q_i$) were extracted using a circle-fitting method \cite{chen2022scattering, baity2024circle}. Measurements were conducted on untreated devices to establish a baseline, and directly compared with ALE+ALD treated devices. 

Our devices exhibit the typical saturation behavior of $Q_i$ as a function of photon number. In the single-photon limit, losses are dominated by unsaturated or partially saturated TLSs, while at higher power levels, $Q_i$ is predominantly limited by other loss mechanisms such as vortices, non-equilibrium quasi-particles, or packaging modes. We quantify the TLS contribution to loss, $\delta_{\text{TLS}}$, from high-power (HP) and low-power (LP) $Q_{i}$ per the following equation \cite{calusine2018analysis}:
\begin{equation}
\delta_{\text{TLS}}=\frac{1}{Q_{\text{TLS}}} = \frac{1}{Q_{i}^{\text{LP}}} - \frac{1}{Q_{i}^{\text{HP}}} 
\end{equation}

Figure \ref{fig:Resonator} shows a power sweep of $\delta_{\text{TLS}}$ for three untreated/treated pairs of representative resonators at 5.3, 5.4 and 5.45 GHz frequencies. At single-photon powers, the loss of the ALE+ALD treated resonators is approximately half that of the untreated resonators. This shows that materials-based losses can be efficiently reduced in single-layer Al devices using the ALE+ALD treatment.

\subsection{Surface Engineering Effects on Transmon Performance}

Motivated by the promising resonator results, we extend our study to transmons made from Al on Si. Here, we compare the transmon qubit energy relaxation time ($T_{1}$), and corresponding effective quality factor ($Q$), before and after the treatment. The measurements were conducted on Josephson junction (JJ) based planar transmons coupled to CPW readout resonators. Five chips, each containing four tunable transmons with variable-size junctions and constant-geometry shunt capacitors ($24$-$\mu$m in-plane dimension, $30$-$\mu$m gap-to-ground plane), were measured in the frequency range of $2.2$–$4.4$~GHz before and after the treatment. Further qubit design details and a micrograph of a representative transmon device are provided in Figure SI-1 of the Supplementary Information (SI). To improve the pre-treatment statistical analysis, six additional untreated chips were measured. One chip was also measured before and after annealing at 300$^{\circ}$C under vacuum in the ALD chamber to disentangle the impact of heating from the ALE+ALD treatment, which also takes place at 300$^{\circ}$C. Transmon and readout resonator frequencies were measured, and qubit $T_{1}$-time data collected over several hours across a broad frequency range. This approach enabled us to sample a wider distribution of possible TLS defects, facilitating faster and more accurate characterization of the $T_{1}$ distribution \cite{barends2013coherent, klimov2018fluctuations}. Based on the standard tunneling model (STM) \cite{clare2022two}, it can be approximated that the TLS density remains constant across the measured frequency range, suggesting a frequency-independent metric of transmon``quality" based on the effective quality factor $Q = 2\pi f_{01} T_1$~ \cite{gambetta2016investigating}, where $f_{01}$ is the ground-to-first excited state transition frequency of the transmon. 

\begin{figure*}[ht!]
    \centering
    \subfloat[] {\includegraphics[width=0.31\textwidth]{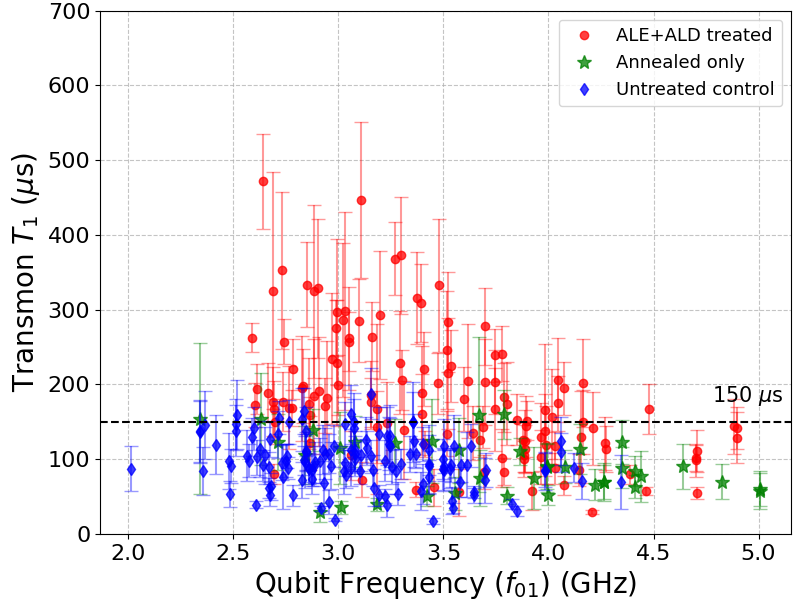}\label{fig:sub1}}\quad%
    \subfloat[] {\includegraphics[width=0.31\textwidth]{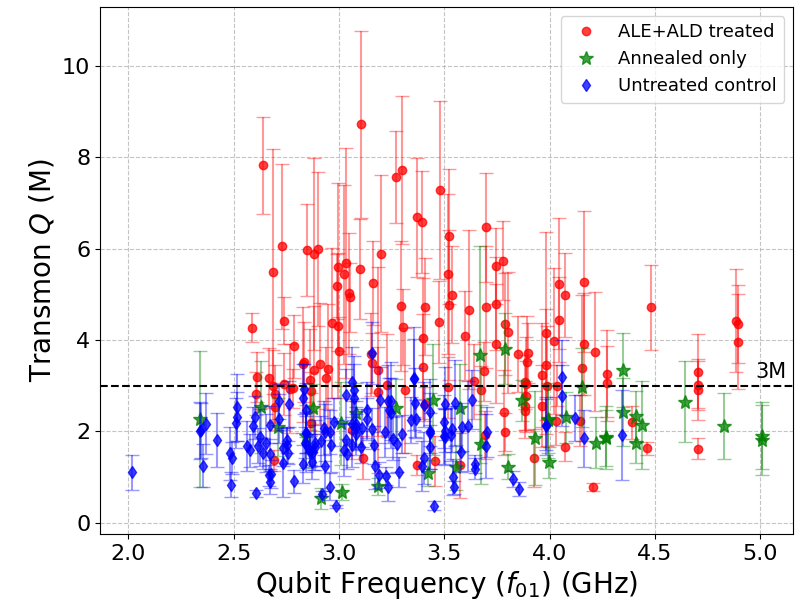}\label{fig:sub2}}\quad%
    \subfloat[]{\includegraphics[width=0.3\textwidth]{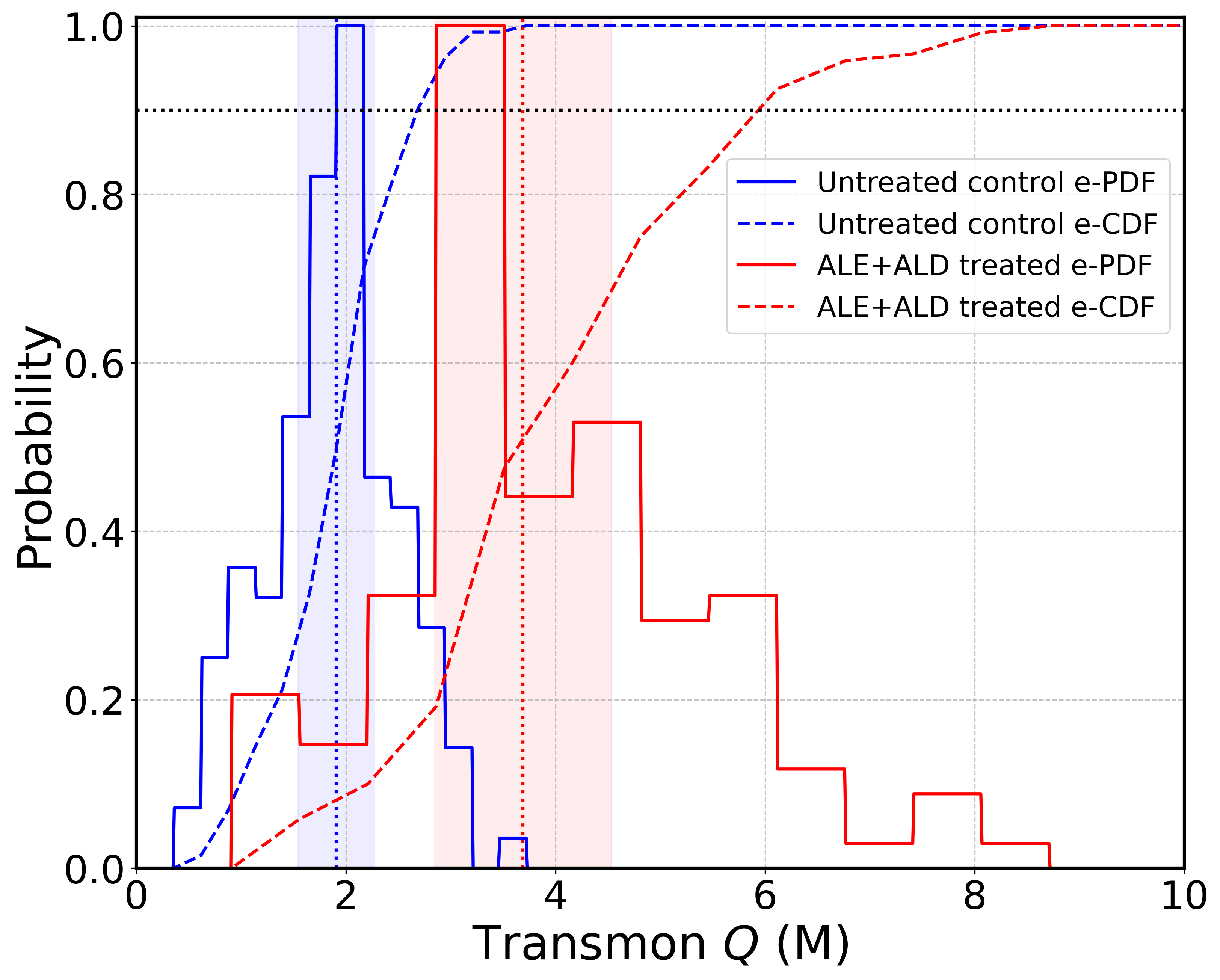}}%
    \caption{Frequency dependence of (a) measured $T_{1}$ times and (b) estimated $Q$ for various devices: untreated, ALE+ALD treated, and annealed-only. The dashed lines are a guide to the eye. (c) The empirical probability distribution function (e-PDF) and cumulative distribution function (e-CDF) are also shown for $Q$ before and after the treatment; the dotted horizontal line, vertical lines and shaded areas represent the 90$^{\text{th}}$-percentile, median, and  median absolute deviation (MAD), respectively.}
    \label{fig:Transmons}
\end{figure*}

$T_1$ and $Q$ measurements of treated and untreated transmons are presented in Figure \ref{fig:Transmons}(a) and (b), respectively. Across the measured frequency range, median $Q$ values remain mainly under $3 \times 10^6$ for the untreated control and annealed-only devices and increase to $3$-$7 \times 10^6$ after the treatment, with some values exceeding the $9 \times 10^6$ mark. Similarly, most median $T_1$ times for the untreated and annealed-only devices are below 150 $\mu$s, whereas treated devices achieve median $T_1$ values in the 150–400 $\mu$s range. These results highlight the effectiveness of the ALE+ALD treatment in reducing the TLS losses in planar transmon devices, consistent with earlier observations from single layer resonators.

This improvement can be further quantified by comparing the pre- and post-treatment empirical probability density function (e-PDF) and cumulative distribution function (e-CDF) of a frequency-independent parameter, $Q$, shown in Figure \ref{fig:Transmons}(c). Prior to the treatment, the e-PDF exhibits a median $Q$ of $1.91 \pm 0.18 \times 10^6$, while the e-CDF shows a 90$^{\text{th}}$-percentile of $2.67 \times 10^6$, which corresponds to a median $T_1$ of $101 \pm 10$~$\mu$s and a 90$^{\text{th}}$-percentile $T_1$ of $142$~$\mu$s for $f_{01}$ of $3$~GHz. Following the treatment, the median $Q$ in the e-PDF increases to $3.69 \pm 0.42 \times 10^6$, while the 90$^{\text{th}}$-percentile $Q$ in the e-CDF rises to $5.9 \times 10^6$, corresponding to a median $T_1$ of $196 \pm 22$~$\mu$s and 90$^{\text{th}}$-percentile $T_1$ of $313$~$\mu$s for $f_{01}$ of $3$~GHz. While this enhancement reflects a reduction in TLS density after treatment, the $T_{1}$ floor remains largely unchanged, likely due to the finite residual population of more strongly coupled TLSs still present on the capacitor pads. As a result, the spread in $T_1$ and $Q$ values increases, leading to a higher standard deviation after treatment. 

The highest performing treated devices are consistently observed within the $2.6$–$3.7$~GHz range, with some of the $Q$ values surpassing $6 \times 10^6$. Purcell loss due to radiation into the qubit read-out resonator, illustrated in Figure SI-2, is presumed to contribute to the reduction in $T_1$ and $Q$ at higher frequencies. Regardless of the Purcell effect, the consistent enhancement in $T_1$ and $Q$ post-ALE+ALD treatment across the studied frequency range, which is not achieved by annealing alone, provides compelling evidence that the surface dielectric loss of the transmons is reduced by the ALE+ALD treatment specifically. No decline in $T_1$ and $Q$ was observed in transmons measured over periods of up to 8-9 months after the treatment, highlighting the long-term stability of this treatment for Al-on-Si based superconducting quantum circuits. 

Additionally, a consistent increase in transmon frequency ($f_{01}$) by several hundreds of MHz was also observed for all treated transmons (as shown in Figure SI-3). This increase originates from junction evolution, as we observe a rise in junction energy (E$_{j}$) after the treatment, while capacitive energy (E$_{c}$) remains unaffected. This frequency shift results from the elevated processing temperatures during treatment, which was also confirmed with annealed-only devices; however, its absolute magnitude is likely influenced by additional factors, such as the initial interface quality before the JJ deposition and fabrication variability. Additionally, we observe that transmon $Q$ decreases with increasing frequency shift (as shown in Figure SI-4), likely due to enhanced Purcell loss as $f_{01}$ approaches the readout frequency.

\subsection{Surface Chemistry}

\begin{figure*}[ht!]
    \centering
    \subfloat[] {\includegraphics[width=0.35\textwidth]{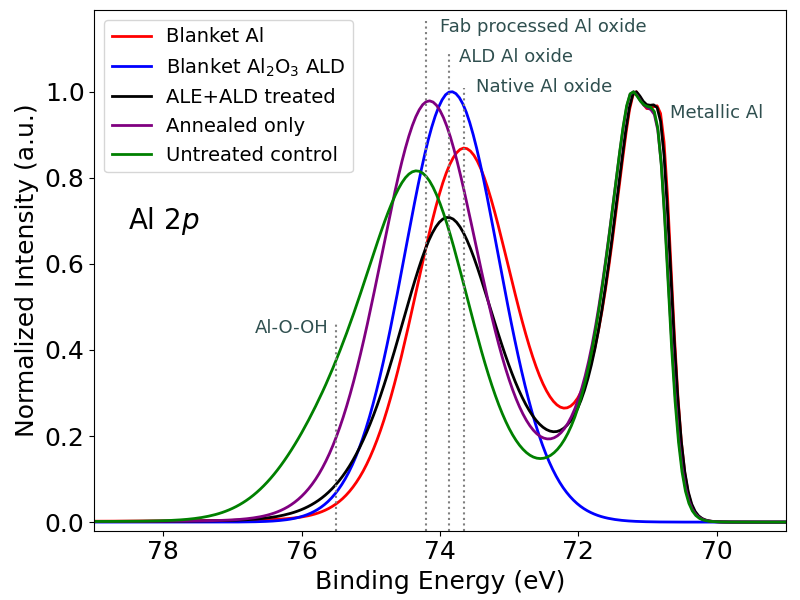}\label{fig:sub3}}\quad%
    \subfloat[] {\includegraphics[width=0.35\textwidth]{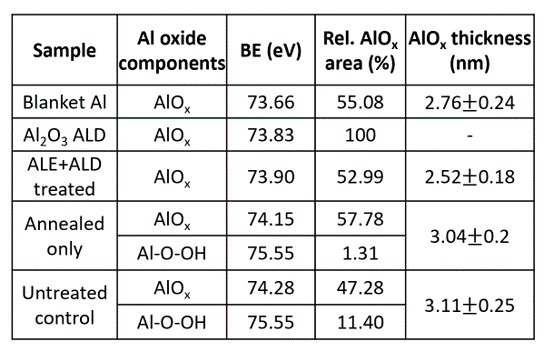}\label{fig:sub4}}\quad%
    \subfloat[] {\includegraphics[width=0.35\textwidth]{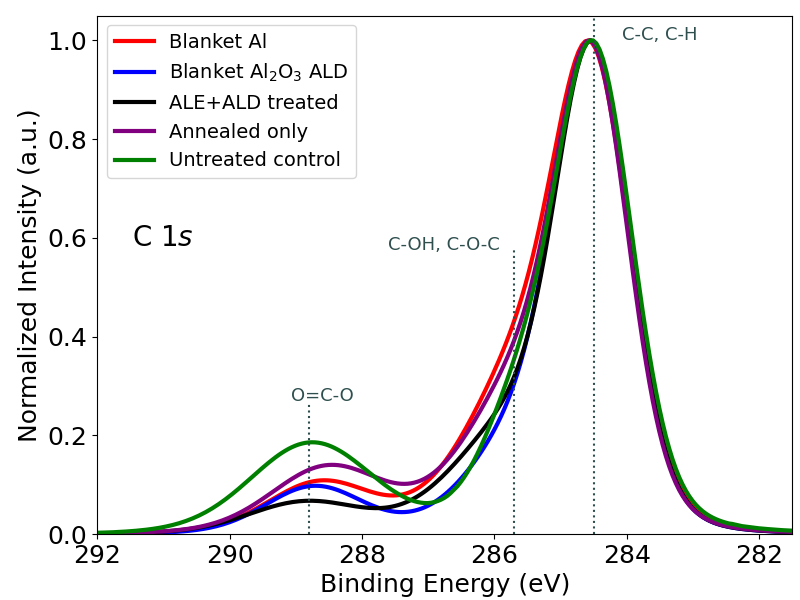}\label{fig:sub5}}\quad%
    \subfloat[]{\includegraphics[width=0.35\textwidth]{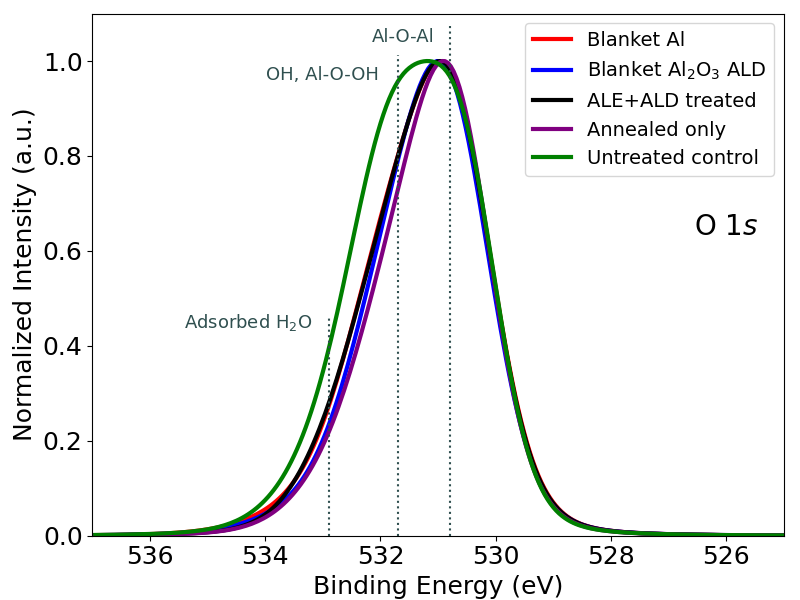}}%
    \caption{(a) Al $2p$, (c) C $1s$, and (d) O $1s$ core-level XPS spectra measured on reference blanket films and transmon qubit chips. The spectra in (a) are aligned in energy and normalized in intensity to the metallic Al $2p_{3/2}$ peak at $\sim$ 70.8 eV, and in (c) and (d) they are calibrated to the C-C bonding feature at 284.5 eV. Dashed lines highlight the different oxidation states and chemical bonds inferred from the relative energy shifts. Information inferred from the deconvolution of the experimental Al $2p$ spectra is summarized in the table shown in (b).}
    \label{fig:XPS1}
\end{figure*}

The ALE chemistry, utilizing alternating exposures of trimethylaluminum (TMA) and hydrogen fluoride-pyridine (HF-pyridine) at 300$^{\circ}$C \cite{lee2016trimethylaluminum}, was developed to etch the native aluminum oxide. The subsequent ALD process, using TMA and water (H$_{2}$O) \cite{puurunen2005surface}, was tuned to regrow a 1-nm thick  Al$_{2}$O$_{3}$ layer. Detailed process parameters and methodology are provided in the Methods section. Here, we characterize the effect of these processes on the chemical properties of the surface of our devices, using a combination of X-ray photoelectron spectroscopy (XPS), photo-induced force infrared microscopy (PiFM), and transmission electron microscopy complemented by electron energy loss spectroscopy (TEM-EELS). Using lab-scale XPS with an Al K$\alpha$ X-ray source (1486.6 eV), we measured the Al $2p$, C $1s$, and O $1s$ core levels on the surface of transmon qubit chips. A detailed description of the fitting procedure is provided in SI (Figures SI-5 through 7). To check the consistency of the observed trends, we measured two sites per untreated and annealed-only chip, and a total of four sites on two different treated chips. 

The XPS data is shown in Figure \ref{fig:XPS1} for the surfaces of untreated, annealed-only, and ALE+ALD treated transmon qubit chips, and compared with blanket films of ALD-grown Al$_2$O$_3$ and evaporated Al. In the fits, up to four characteristic binding energies (BE) can be distinguished for the Al $2p$ oxide components. Ranked by increasing BE (see Figure \ref{fig:XPS1}(a)), these are assigned to native Al oxide, ALD-grown Al$_{2}$O$_{3}$, post-fabrication AlO$_{x}$, and an aluminum oxyhydroxide phase (Al-O-OH). The differences in BE (see table in Figure \ref{fig:XPS1}(b)) were found to be consistent across the measured chips and sites. The higher BE of the oxide components for the untreated and annealed-only devices suggests the formation of an oxygen-rich aluminum oxide on their surface \cite{prasanna2013composition, reddy2014xps, mengesha2020corrosion}. Additionally, the untreated chip shows a relatively strong Al-O-OH component, which is commonly found on surfaces subjected to wet chemical treatments. Annealing removes most of those Al-O-OH groups, while ALE+ALD removes all of them. Furthermore, we find that the ALE+ALD-treated Al surface has the thinnest oxide layer ($2.52 \pm 0.18$ nm), followed by the native oxide of the Al blanket film ($2.76 \pm 0.24$ nm), and then the untreated and annealed-only ($3.1 \pm 0.25$ nm) chips. The thinning of the surface oxide layer of Al after ALE+ALD was confirmed by TEM-EELS, see Figure SI-8(a-d). The fact that the surface oxide in the ALE+ALD treated chip is thicker than the anticipated 1-nm thick ALD-grown layer can be attributed to the partial regrowth of a native oxide layer during the 20 seconds that the sample spends at 300$^{\circ}$C in the 6-7 Torr atmosphere of the process chamber upon completion of the ALE step and prior to the start of the ALD process \cite{nguyen2018atomic}.

The deconvolution of the C $1s$ core level reveals three distinct chemical environments of carbon, which, in order of increasing BE, are: adventitious carbon or C-C bonding, C-OH or C-O-C bonding, and O=C-O bonding. The reduced intensity of the O=C-O bonding component after annealing, but especially after ALE+ALD where it is even weaker than in blanket Al, points to a reduction of organic species primarily stemming from residual resist contamination. The clear broadening of the O $1s$ peak observed in the untreated chip surface can be assigned to a contribution from adsorbed hydroxyl species such as Al-O-OH or Al(OH)$_{3}$ around 531.7 eV and adsorbed H$_2$O near 533 eV. These features are not observed after annealing or ALE+ALD treatment, which is consistent with the Al $2p$ data. Taken together, the XPS data shows that the ALE+ALD treatment results in a thinner and stoichiometric surface layer of Al$_{2}$O$_{3}$ with minimal polymeric contamination.

\begin{figure}[htbp]
    \centering
    \includegraphics[width=0.48\textwidth]{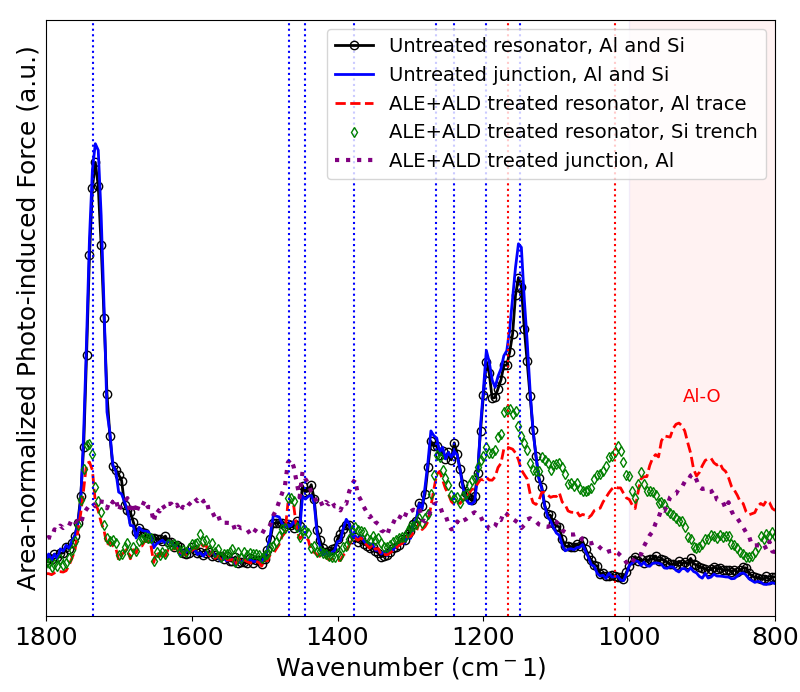}
    \caption{Comparison of area-normalized PiFM spectra averaged across Si and Al regions measured on the untreated and ALE+ALD-treated transmon qubit chips. Dashed lines indicate characteristic peaks: blue for those present before and after the treatment, red for peaks that emerged after the treatment, and red shaded area represents the broad Al-O stretching vibration appearing only after the ALE+ALD treatment.}
    \label{fig:PiFM}
\end{figure}

In order to gain spatially-resolved insights into this surface modification, we used PiFM to collect infrared spectroscopic data in the 800-1800 cm$^{-1}$ range with a $\sim$10 nm lateral resolution. Areas of a few $\mu$m$^{2}$ centered around a junction and a portion of the Al/Si sidewall from a readout resonator were measured on untreated and ALE+ALD treated qubit chips. To introduce the spectroscopic features detected on the surface of our samples, Figure \ref{fig:PiFM} shows the PiFM spectra averaged over the Si and Al regions, with each spectrum obtained by averaging 20-30 individual scans across the measured areas. The spectra are area-normalized to account for differences in absolute intensity and enable a more direct comparison of spectral features between the two samples. Dashed lines serve as visual guides for the wavenumbers of peaks present either before and after the treatment (blue) or only after the treatment (red), respectively. A detailed peak assignment is provided in Table 1 in the SI. In the untreated chip, the signal measured across all areas is dominated by spectral features assigned to poly(methyl methacrylate) (PMMA, blue dashed lines in Figure \ref{fig:PiFM}) \cite{duan2008preparation, aziz2019effect}, which is a resist used during liftoff and dicing steps of our fabrication flow. Despite soaking the chips in a heated solution of N-methyl pyrrolidone (NMP) followed by an acetone and isopropyl alcohol rinse to strip the PMMA, the absence of a signal from AlO$_x$ suggests that sufficient residual PMMA remains to form a continuous layer of at least $\sim$2 nm thickness on the untreated surface. This lower bound on the residual PMMA thickness is derived from tests conducted to estimate the minimum thickness of a thin polymer layer that can effectively shield the infrared signal from the underlying sample.

Conversely, we observe a broad absorbance feature around 850-1000 cm$^{-1}$ assigned to Al-O stretching vibrations \cite{goldstein2008al2o3,vemuri2023comprehensive} on both the resonator and junction areas of the ALE+ALD treated chip as shown by the shaded area in Figure \ref{fig:PiFM}. A peak corresponding to C-O stretching vibrations also appears at 1020 cm$^{-1}$ (red dashed line) in the resonator area, suggesting PMMA decomposition. Previous studies showed that, at elevated temperatures, PMMA degrades into its monomer, MMA, which shares similar infrared features and can further undergo random scission around 300$^{\circ}$C, breaking ester bonds into smaller fragments like C–O \cite{kashiwagi1989behavior, manring1991thermal}. Additionally, we find that the characteristic PMMA absorbance features are weaker in the resonator area and nearly suppressed in the junction area, indicating a significant reduction in the overall PMMA contamination. The fact that the junction surface is cleaner can be attributed to the fewer exposures of the junction to MMA- and PMMA-containing resists compared to the ground plane, resulting in a thinner contamination layer and a more effective removal. Alongside these changes, a new peak appears near 1170 cm$^{-1}$ (red dashed line), attributed to C-H bending vibrations from Al-CH$_3$ bonds. Such features have been previously reported during the ALD or ALE reactions involving TMA as one of the precursors \cite{al2003ft, goldstein2008al2o3, lee2015atomic, dumont2017competition}. Additionally, the selectivity of the Al$_2$O$_3$ growth in the nucleation regime is highly dependent on the functionalization of the starting surface, particularly for Si \cite{gakis2019iotan, xu2022area, xu2022extending}. This can explain the absence of a feature corresponding to Al-O stretching in the 850-1000 cm$^{-1}$ range in the Si trench area, which suggests that no Al$_2$O$_3$ is formed on the Si surface. The absence of Al$_2$O$_3$ on the Si surface following the ALE+ALD treatment is further confirmed by the TEM-EELS maps shown in Figure SI-8(e). 

The effectiveness of the ALE+ALD treatment in reducing PMMA contamination on both Al and Si surfaces despite the intended selectivity of our ALE chemistry for AlO$_x$ may seem surprising. The strong adhesion of PMMA to aluminum and silicon oxides has been previously attributed to the bonding between PMMA and the surface hydroxyl groups \cite{papirer1994adsorption, watts2000adsorption}. The annealing-induced removal of hydroxyl bonds, suggested by the Al $2p$ and O $1s$ XPS data for the annealed-only chips, is expected to substantially weaken the adhesion of PMMA to both Al and Si surfaces, making it easier for the HF used in the ALE chemistry to indiscriminately etch the PMMA away. The degradation of PMMA at the ALE process temperature (300$^{\circ}$C) \cite{ferriol2003thermal}, as described above, is also anticipated to ease the etch.

\begin{figure}
    \centering
    \includegraphics[width=0.48\textwidth]{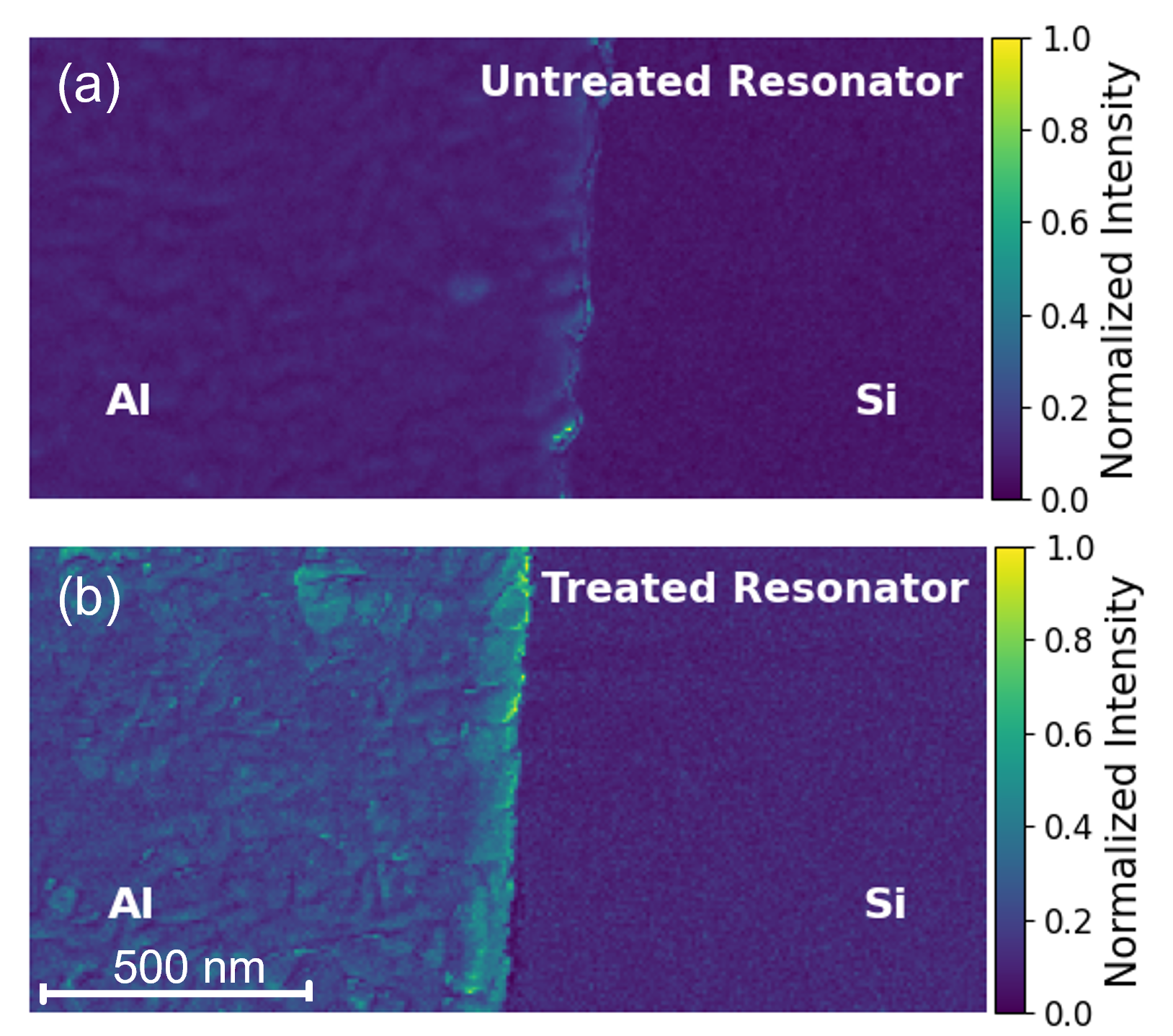}%
    \caption{(a) and (b) PiFM maps measured at 912 cm$^{-1}$ (characteristic of Al-O) normalized by the map measured at 1740 cm$^{-1}$ (characteristic of PMMA), for untreated and treated resonator areas, respectively.}
    \label{fig:PiFM maps}
\end{figure}

 Chemical maps were measured to gain detailed insight into the spatial uniformity of our surface treatment. Figures \ref{fig:PiFM maps}(a) and (b) present PiFM maps for untreated and treated resonator area, respectively. They were measured at 912 cm$^{-1}$, corresponding to the Al–O stretching vibration, and normalized by the map measured at 1740 cm$^{-1}$, associated with the C=O stretching vibration in PMMA. This normalization reduces local variations of the gap-field enhancement from topographic effects \cite{almajhadi2017contrast}, which are considered to be constant across the measured spectral range, thereby enhancing imaging contrast based on the local ratio of Al oxide to PMMA. The raw maps are shown in Figure SI-9. The signal at 912 cm$^{-1}$ is barely visible in the map of the untreated surface, consistent with a uniform coverage of the aluminum oxide by PMMA. After treatment, the surface shows a stronger Al-O signal across the entire Al surface. This indicates a uniform thinning of the PMMA residues revealing the aluminum oxide underneath.

\subsection{Discussion}

We showed that post-fabrication ALE+ALD can improve resonator and transmon performance by treating both top and sidewall surfaces, and provide compatibility with Al unlike conventional wet chemical treatments like buffered oxide etchant (BOE) or triacid \cite{altoe2022localization, crowley2023disentangling, mclellan2023chemical}. We find that ALE+ALD treated surfaces exhibit a reduction in polymeric residues, along with a more stoichiometric Al native oxide. While our study does not disentangle the individual contributions of polymeric residue removal and Al oxide layer modification, it establishes a direct correlation between improved material quality at the MA and SA interfaces and increased resonator $Q$ and transmon $T_{1}$. The comparable improvement observed in both the resonator and transmon $Q$ further suggests that the weakly coupled TLS defects located on the surfaces of the capacitor pads limit the lifetime of our transmon qubits.

Our findings also suggest pathways for future optimization of our process. These include incorporating a step in the ALE process to remove the silicon oxide from the substrate surface, gaining a deeper understanding of the mechanism by which ALE etches organic contamination to further enhance process efficiency, and exploring lower-loss dielectric coatings beyond Al$_{2}$O$_{3}$. Finally, we find that changes in the junction also occur during the ALE/ALD process, which can lead to large shifts in the junction resistance; further studies are necessary to understand both the mechanism for, or source of, the resistance change, and the intrinsic variability of the change.

\section{Conclusion}

We have demonstrated that the energy-relaxation times of Al transmons can be systematically and durably increased by more than two-fold using a combination of ALE and ALD. These processes uniquely enable post-fabrication surface treatment by removing fabrication residues from all exposed surfaces and replacing the Al native oxide from the top surface and sidewalls with a conformal, stoichiometric Al$_{2}$O$_{3}$ layer. Another differentiator of this approach from other surface treatments is the full compatibility with Al, a rarity among commonly used native oxide etchants. The concurrent doubling of the quality factor in Al-based resonators and transmons subjected to the same ALE+ALD treatment suggests that the observed improvement primarily stems from a reduction in the density of coherence-limiting TLS defects located on the surface of the capacitor pads. These findings establish a clear correlation between TLS loss, transmon relaxation times, and the chemical composition and purity of the MA and SA interfaces. They also provide a framework for continued improvements in quantum device performance through controlled, conformal, and Al-compatible surface treatments.

\section{Methods}
\subsection{Surface Treatment: ALE and ALD process}
\begin{figure*}[t]
    \centering
    \subfloat[] 
    {\includegraphics[width=0.48\textwidth]{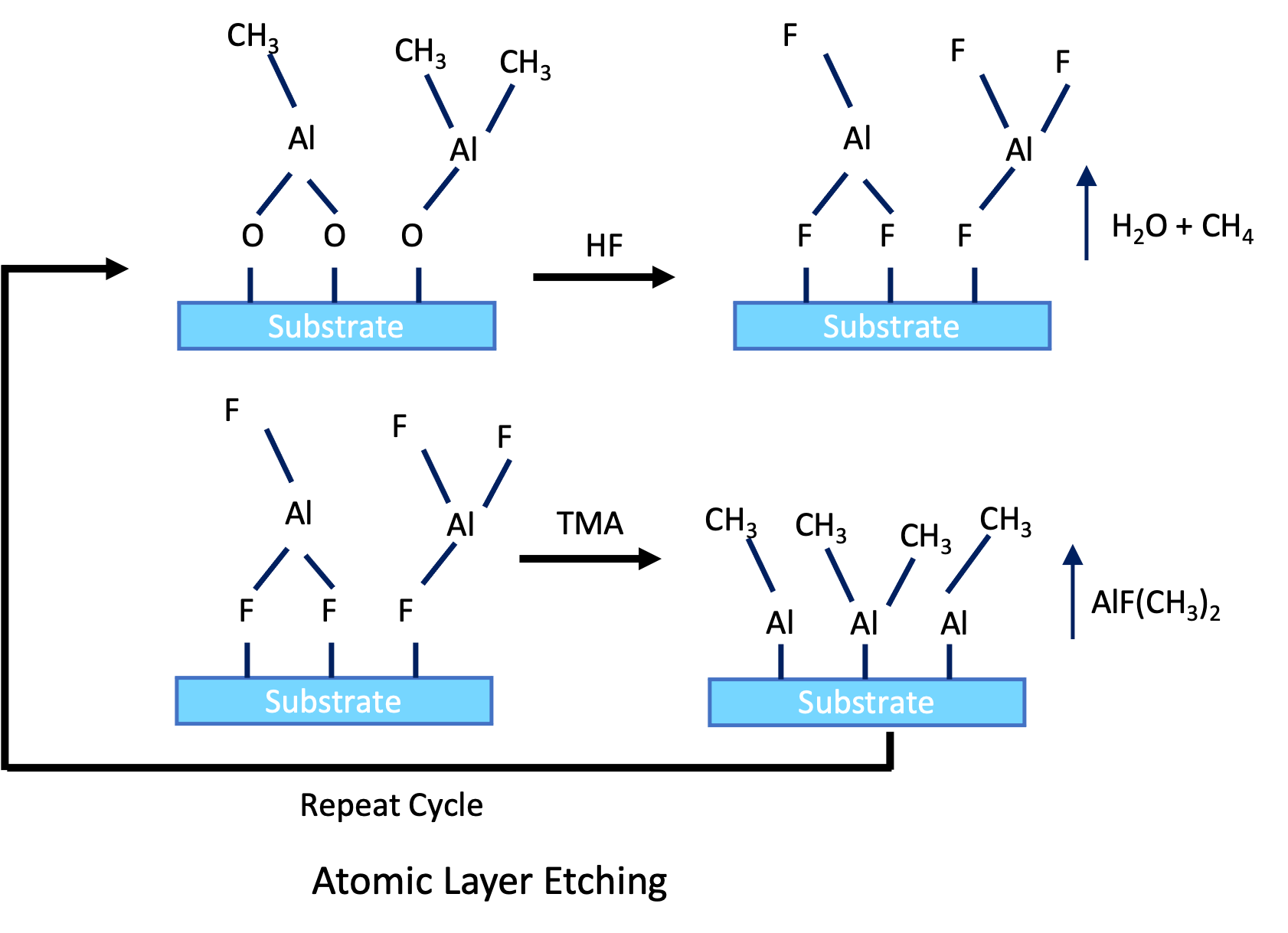}\label{fig:sub24}}
    \hspace{0.02\textwidth} 
    \subfloat[] 
    {\includegraphics[width=0.48\textwidth]{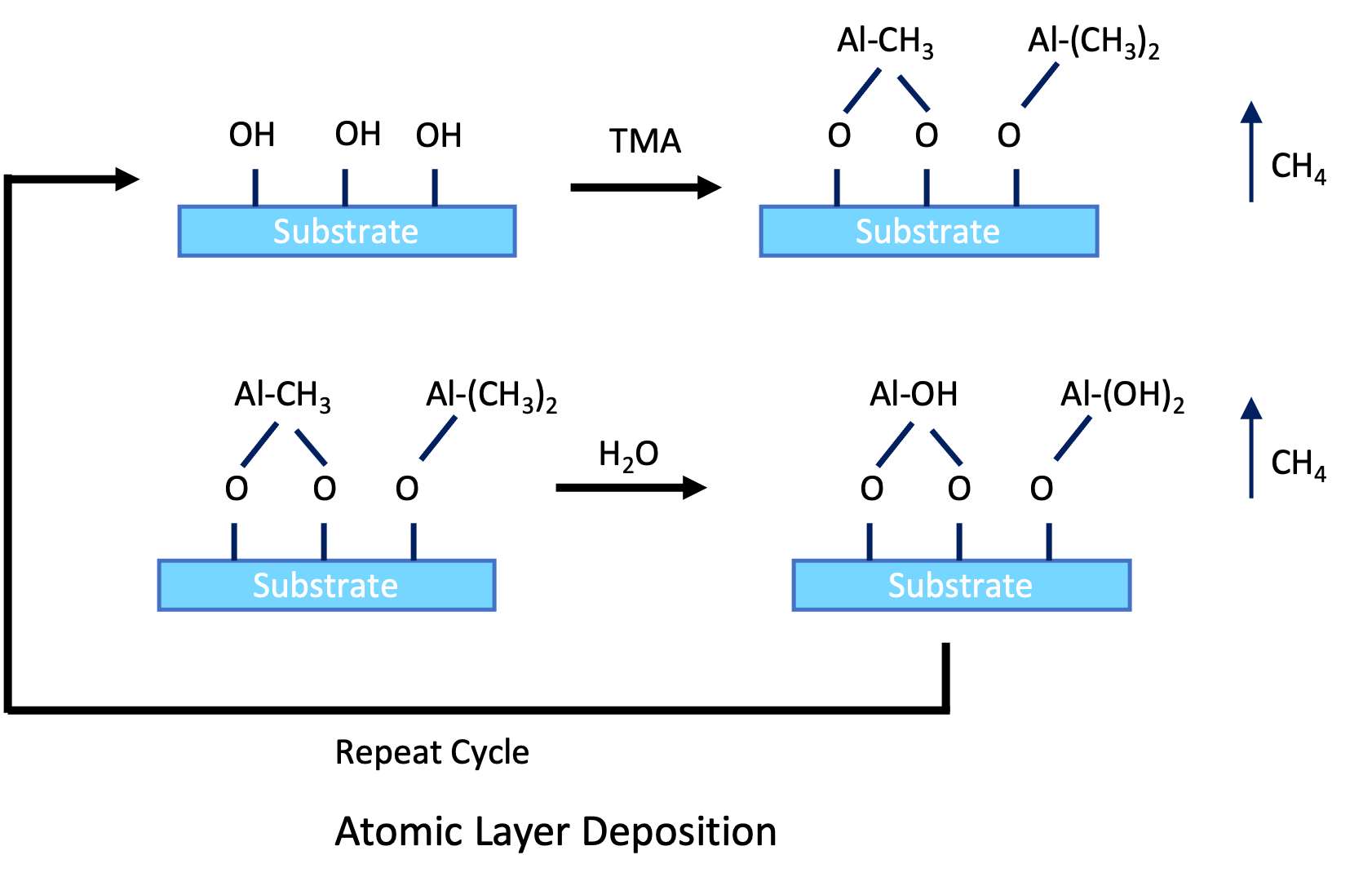}\label{fig:sub25}}
    \caption{Schematic representation of (a) ALE and (b) ALD using self-limiting surface chemistry and an AB binary reaction sequence.}
    \label{fig:ALE+ALD schematic}
\end{figure*}

The ALE chemistry developed for this work relies on alternating exposures of HF-pyridine and TMA at 300$^{\circ}$C resulting in an etch rate of 0.5 \AA/cycle \cite{lee2016trimethylaluminum}. The reaction mechanism for the ALE of Al$_{2}$O$_{3}$ is depicted in Figure \ref{fig:ALE+ALD schematic}(a). The proposed two-half cycle reaction can be written as follows: 
\begin{equation}
\mathrm{Al_2O_3^* + 6HF (g) \rightarrow 2AlF_3^* + 3H_2O(g) \uparrow}
\end{equation}
\begin{equation}
\mathrm{2AlF_3^* + 4Al(CH_3)_3 (g) \rightarrow 6AlF(CH_3)_2 (g) \uparrow} 
\end{equation}
where the asterisk (*) denotes the surface species at the end of each half-cycle. In this ALE chemistry, HF reacts with the top few monolayers of Al$_{2}$O$_{3}$ to form AlF$_{3}$ in the first half cycle `A' (equation (2)). In the second half cycles `B' (equation (3)), AlF$_{3}$ is subsequently removed from the surface as AlF(CH$_{3}$)$_{2}$, a volatile species at 300$^{\circ}$C, when reacted with TMA. Repetition of ABAB.. half cycles leads to the layer-by-layer etching. 

Following the successful removal of the native Al oxide, the surface is encapsulated \textit{in situ} with a thin layer of Al${_2}$O${_3}$ via ALD. The Al${_2}$O${_3}$ ALD process involves alternating exposures of TMA and water (H${_2}$O) \cite{puurunen2005surface} at the same temperature of 300$^{\circ}$C, resulting in a growth rate of 1 \AA/cycle. This process begins immediately after the ALE to minimize spontaneous re-oxidation of the underlying Al surface. The reaction mechanism for the ALD of Al$_{2}$O$_{3}$ is depicted in Figure \ref{fig:ALE+ALD schematic}(b). The proposed two-half cycle reaction can be written as follows: 
\begin{equation}
\mathrm{AlOH^* + Al(CH_3)_3 (g)\rightarrow AlOAl(CH_3)_2^* + CH_4(g)\uparrow}   
\end{equation}
\begin{equation}
\mathrm{AlOAl(CH_3)_2^* + H_2O (g) \rightarrow AlOH^* + CH_4(g) \uparrow}  
\end{equation}
The ALD surface chemistry involves formation of Al-O-Al(CH${_3}$)${_2}$ surface species during the first half cycle `C' (equation (4)) when Al-OH* surface species react with TMA. In the second half cycle `D' (equation (5)), these species are converted back to Al-OH* when reacted with H${_2}$O, ultimately leading to the formation of the Al${_2}$O${_3}$ monolayer on the surface.

The ALD and ALE processes were developed using a Picosun/AMAT R-200 tool at 300$^{\circ}$C. TMA (99.9999\% purity from Strem) and HPLC grade water (H$_{2}$O) were used as the metal and the oxygen precursors for the Al$_{2}$O$_{3}$ ALD process while TMA and HF-pyridine (70\% from Sigma Aldrich) served as the metal and fluorine sources for the etching of Al$_{2}$O$_{3}$. All the precursors were kept at room temperature. High purity nitrogen (N$_{2}$, 99.999\% purity) was employed as the carrier gas to carry the precursors into the reaction chamber. The process base pressure was maintained at $\sim$ 11 hPa, with 300 sccm of N$_{2}$ directed into the chamber and $\sim$ 50 sccm of N$_{2}$ through all the precursor lines. The partial pressures of TMA, Al$_{2}$O$_{3}$ and HF-pyridine were maintained at 1.7 \(\pm\) 0.2 hPa, 2.3 \(\pm\) 0.2 hPa and 4 \(\pm\) 0.5 hPa, respectively. N$_{2}$ was also used as a purge gas between successive doses of reactants. The optimized dosing sequence can be expressed as (t$_{1}$ - t$_{2}$ - t$_{3}$ - t$_{4}$), where t$_{1}$ is the dose time of TMA and t$_{3}$ is the dose time of H$_{2}$O or HF-pyridine, with t$_{2}$ and t$_{4}$ representing intermediate purge times. All the times are presented in seconds. 
The optimized dosing sequence for ALD Al$_{2}$O$_{3}$ was (0.2-10-0.2-10), while the dosing sequence for ALE Al$_{2}$O$_{3}$ was (1-10-1-10). The combined ALE+ALD process includes 50 cycles of Al$_{2}$O$_{3}$ ALE, followed by 10 cycles of Al$_{2}$O$_{3}$ ALD at 300$^{\circ}$C. The combined recipe was applied to individual resonator and transmon chips, enabling the \textit{in situ} removal of native Al oxide followed by encapsulation with $\sim$ 1 nm of ALD Al$_{2}$O$_{3}$ without breaking vacuum. 

\subsection{Device Fabrication}

The 2D transmons and resonators were fabricated on high-resistivity silicon ($>20\,\mathrm{k\Omega\cdot cm}$) wafers, beginning with a 10-minute remote oxygen plasma clean at 150 W in a Tergeo asher. The first step was to deposit the Al ground layer using a lift-off process, which began with spin coating an adhesion promoter (SurPass 3000) and resist (ma-N 2403), followed by baking at 90$^{\circ}$C, electron beam (e-beam) lithography, and development using ma-D 525. The wafers were then treated with a 10-second direct oxygen plasma exposure in an Oxford etcher and a 15-second immersion in 10:1 BOE, prior to depositing a 100-nm thick Al layer with a 10 \AA/s growth rate under a base pressure of 10$^{-9} $ Torr in an Angstrom e-beam evaporator. The lift-off process was completed by soaking the wafers in a hot NMP solution at 150$^{\circ}$C for 2 hours, followed by 5 minutes of sonication in acetone and thorough rinsing with IPA and acetone.

For the second layer (JJ), the Al ground plane was first exposed to a 5-minute remote oxygen plasma treatment in a Tergeo asher. This was followed by spin coating bi-layer resists (MMA EL11 and PMMA A4 950K), with bake steps at 170$^{\circ}$C both before and after spin coating. After e-beam lithography and cold development in a 3:1 IPA:H$_{2}$O solution at 1.6$^{\circ}$C, the wafers underwent an O$_{2}$ plasma descum step in an Oxford etcher. They were then exposed to a 5-minute vapor HF treatment and loaded into an Angstrom e-beam double angle evaporator within 5 minutes for junction deposition (same growth rate and base pressure as the ground layer), with deposition angles of 60$^{\circ}$ and 0$^{\circ}$. Finally, the resist was removed with a 2-hour hot NMP treatment at 150$^{\circ}$C, followed by 5 minutes of sonication in acetone, thorough rinsing with IPA and acetone, and a 10-minute remote oxygen plasma treatment in a Tergeo asher to remove any residual photoresist.

Scaffold oxide features were fabricated in the next step by exposing the wafers to a 10-minute remote oxygen plasma treatment in a Tergeo asher, followed by spin coating of a tri-layer resist stack (PMMA A4 950K, PMGI SF11 and ZEP 520a) with pre- and post-spin coating bake steps at 182$^{\circ}$C. After e-beam lithography, the wafers were developed sequentially in three solutions: ZED-N50 to develop the ZEP 520a layer, MF-319 for the PMGI SF11 layer, and 3:1 IPA:H$_{2}$O mixture for the PMMA A4 layer, followed by an O$_{2}$ descum step in an Oxford etcher. A 1 $\mu$m SiO$_{2}$ layer was deposited using an Angstrom evaporator at a growth rate of 1 \AA/sec and base pressure below 10$^{-6}$ Torr. The oxide layer was lifted off in a 2-hour hot NMP treatment at 150$^{\circ}$C, followed by rinsing with IPA and acetone and a final 10-minute remote oxygen plasma treatment in a Tergeo asher. For the Al airbridges, which form the final fabrication layer, the same process was used except that the SiO$_{2}$ deposition was replaced by \textit{in situ} pre-deposition ion milling and deposition of a 400-nm thick Al layer in an Angstrom e-beam evaporator at a 10 \AA/s growth rate and a base pressure of 10$^{-9} $ Torr.

As a final step, the wafers underwent a 10-minute remote oxygen plasma treatment in a Tergeo asher, followed by spin coating of PMMA EL11 resist and baking at 182$^{\circ}$C. The wafers were then diced using an ADT dicing saw, UV-cured for 7 minutes, and subsequently singulated and sorted. After dicing, the resist was removed with a 10-minute hot NMP bath at 150$^{\circ}$C, followed by rinsing with IPA and acetone and a 10-minute remote oxygen plasma treatment in a Tergeo asher. Vapor HF treatment was used to remove the scaffold oxide, thereby releasing the airbridges and completing the fabrication. The devices were then either subjected to ALE+ALD treatment or left untreated as control samples before mounting on the PCB for wirebonding.

\subsection{Resonator and Transmon Measurement}

Measurements were conducted using a vector network analyzer (VNA) in a conventional set up to capture the $S_{21}$ transmission of the resonators in a dilution refrigerator (DR) with a base temperature of 10 mK. The input lines consists of 75 dB total input attenuation and output line consists of a high electron mobility transistor (HEMT) amplifier at 4K stage and a room temperature amplifier. 

To measure the resonator internal loss, we implemented a parameter-constrained diameter correction method (PC-DCM) \cite{chen2025efficient}, which reduces the parameter space in resonator data fitting and the uncertainty in the fitting result. Key parameters such as the coupling quality factor ($Q_{c}$) and impedance mismatch angle ($\phi$) are extracted from high-power measurements and fixed for low-power analysis, improving fitting efficiency and robustness. For low-power measurements, we reduced the $S_{21}$ frequency span to within one line-width of the resonance.

The quality factors reported in this study are based on measurements across three resonators on two devices for both untreated control and ALE+ALD-treated samples. The TLS-induced loss (\(\delta_{\text{TLS}}=\frac{1}{Q_{\text{TLS}}})\) is derived from the quality factors measured at low power (\({1}/{Q_{i,\text{LP}}}\), collected at 1 photon or less) and high power (\({1}/{Q_{i,\text{HP}}}\)) as \(\delta_{\text{TLS}}=\frac{1}{Q_{\text{TLS}}} = \frac{1}{Q_{i,\text{LP}}} - \frac{1}{Q_{i,\text{HP}}}\).\newline

Similar to the resonator measurements, transmon measurements were conducted using the same DR setup at a base temperature of 10 mK. Further details on packaging and the cryogenic setup for transmon measurements can be found elsewhere \cite{levine2024demonstrating}. $T_1$ and $T_2$ were estimated by fitting an exponential decay to the inversion recovery \cite{wei2020measure, montanez2022decoherence} and an exponentially decaying sinusoidal to the Ramsey curve \cite{montanez2022decoherence}, respectively, with each operating point measured for at least 3 hours. The measured anharmonicity was within 10\% of the design value, which is approximately 220 MHz for all transmons. Notably, the anharmonicity remained unchanged after both ALE+ALD treatment and annealing. We measured transmon decoherence at various flux points away from the sweet spot (zero flux) over a frequency range of 1.0 to 1.5 GHz from the design value. While the designed qubit frequencies were initially between 4 and 5.5 GHz, factors such as junction energy mis-targeting due to fabrication variabilities, high-temperature annealing effects, and $T_1$ measurements taken off the sweet spot shifted the qubit frequencies to 2.5–5 GHz.

\subsection{Materials Characterization}

The XPS measurements were performed using a Scienta Omicron instrument, with a 300 W Al K$\alpha$ X-ray source (1486.6 eV) and a base pressure of 10$^{-10} $ mbar. The analysis area was $\sim$ 400\,$\mu\text{m}$, and photoelectrons were collected at normal take-off angle. The acquired spectra were energy calibrated using the C 1s peak at $\sim$ 284.5 eV.

PiFM measurements were performed in ambient conditions using a VistaScope (Molecular Vista Inc.) system, coupled with a LaserTune QCL. The system covered an infrared spectral range from 760 to 1800 cm$^{-1}$ with a spectral resolution of 4 cm$^{-1}$.

The morphology, surface oxide thickness, and elemental distribution at the Al and Si interfaces in the resonator and junction areas of our transmon devices, both untreated and ALE+ALD treated, were imaged using STEM-EELS. Samples were prepared via a focused ion beam (FIB) lift-out technique, using spin-on polymer (SOP) and electron-beam deposited platinum (e-Pt) before milling. The lamella thickness was $\sim$ 70 nm. Measurements were performed on an FEI probe-corrected TITAN operated at 200 kV, in bright-field STEM and Z-contrast high-angle annular dark-field (HAADF) STEM modes, achieving a resolution of $\sim$ 0.78 Å.

\section{Acknowledgements}

We thank the staff from across the AWS Center for Quantum Computing (CQC) that enabled this project. We also thank Simone Severini and Peter DeSantis for their support of the research activities at the CQC.

\bibliographystyle{ieeetr}

\bibliography{References}

\clearpage

\onecolumngrid
\renewcommand{\thefigure}{SI-\arabic{figure}}
\setcounter{figure}{0}  
\setcounter{section}{0}

\newpage
\section*{Supplementary Information for ``Improving the lifetime of aluminum-based superconducting qubits through atomic layer etching and deposition"}

\input{Supplementary_Information}  

\end{document}

%% file: Supplementary_Information.tex

\section*{Outline}

\begin{enumerate}[label=\Roman*.]
    \item Transmon design and measurement parameters
    \item Calculated Purcell curves and measured $Q$ for individual transmons
    \item Frequency shift after ALE+ALD treatment for individual transmons
    \item Transmon $Q$ dependence as a function of frequency shift after ALE+ALD treatment
    \item XPS peak fitting
    \item Estimation of Al$_2$O$_3$ thickness from XPS analysis
    \item TEM-EELS analysis
    \item IR peak assignment
    \item PiFM mapping of resonator area
\end{enumerate}
\newpage

    
\section{Transmon design and measurement parameters}

\begin{figure}[h]
    \centering
   \subfloat[] {\includegraphics[width=0.2\textwidth]{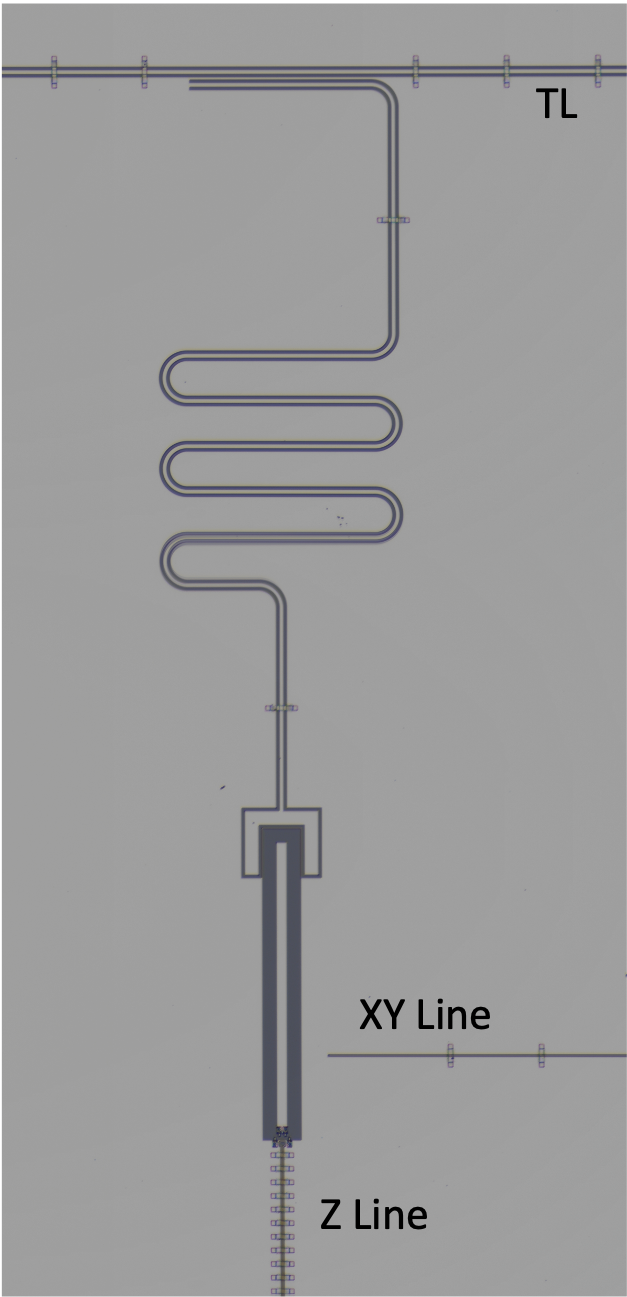}\label{fig:sub6}}\quad%
   \subfloat[]{
    \resizebox{0.65\textwidth}{!}{
    \begin{tabular}{|c|c|c|c|c|}
    \hline
    \textbf{Parameter} & \textbf{Q1} & \textbf{Q2} & \textbf{Q3} & \textbf{Q4} \\ 
    \hline
    Measurement $f_{01}$ range (GHz) & 2.3-3.9 & 2.6-4.4 & 2.4-4.3 & 2.9-4.9 \\
    \hline
    Readout resonator frequency (GHz) & 5.55 & 6.4 & 5.8 & 6.56 \\
    \hline
    JJ length (nm)  & 200 & 200 & 200 & 200 \\
    \hline
    JJ width (nm) & 163 & 200 & 157 & 286 \\
    \hline
    Capacitor width (um) & 24 & 24 & 24 & 24  \\
    \hline
    Capacitor gap (um) & 30 & 30 & 30 & 30 \\
    \hline
    Capacitor length (um) & 684.2 & 684.2 & 645 & 631 \\
    \hline
    Coupling strength (\textit{g}) (MHz) & 59.7 & 78.4 & 60.6 & 61.1 \\
    \hline
    Coupling capacitance ($C_{g}$) (fF) & 5.6 & 6.23 & 5.27 & 4.32 \\
    \hline
    Qubit capacitance ($C_{q}$) (fF) & 87.1 & 87.1 & 86.3 & 84.7 \\
    \hline
    \end{tabular}
    }
    \label{tab:qubit_design}}
    \caption{(a) Optical micrograph of a transmon device and (b) summary of transmon design and measurement parameters.}
    \label{fig:transmon_combined}
\end{figure}

Five different chips, each containing four tunable transmons (Q1-Q4), were measured in the frequency range of 2.2–4.4 GHz before and after the treatment along with six additional untreated chips. One chip was also measured before and after annealing at 300$^{\circ}$C under vacuum in the ALD chamber. All these chips came from two different wafers fabricated with the same process flow. A micrograph of a transmon is shown in Figure \ref{fig:transmon_combined}(a). Design and measurement parameters are summarized in the table in Figure \ref{fig:transmon_combined}(b). 

\section{Calculated Purcell curves and measured $Q$ for individual transmons}

\begin{figure}[htbp]
    \centering
    \subfloat[] {\includegraphics[width=0.45\textwidth]{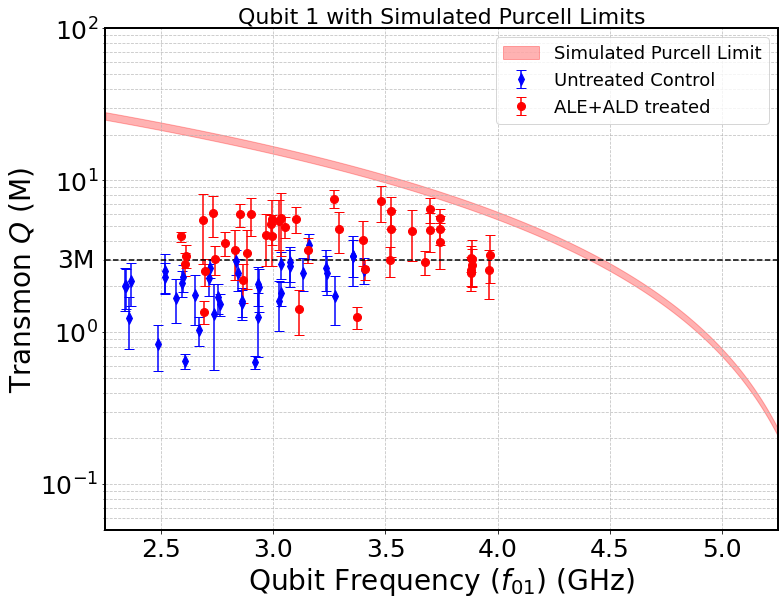}\label{fig:sub7}}\quad%
    \subfloat[] {\includegraphics[width=0.45\textwidth]{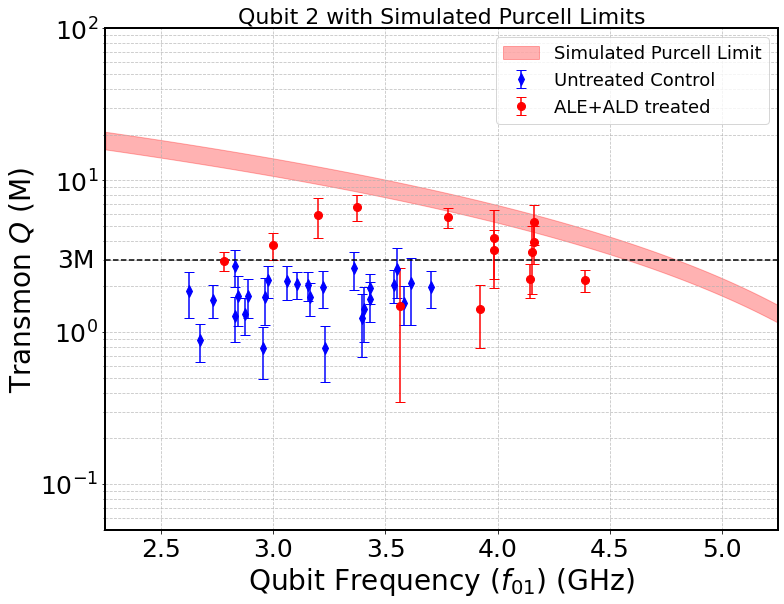}\label{fig:sub8}}\quad%
    \subfloat[] {\includegraphics[width=0.45\textwidth]{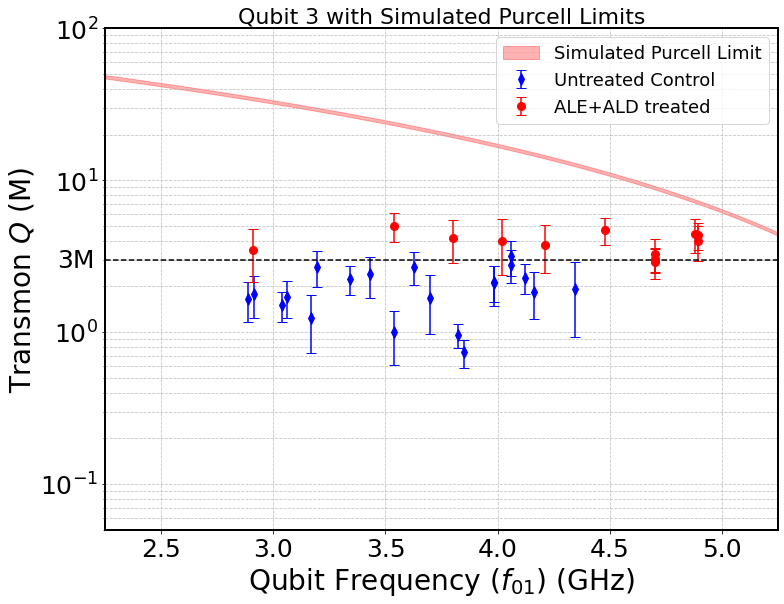}\label{fig:sub9}}\quad%
    \subfloat[]{\includegraphics[width=0.45\textwidth]{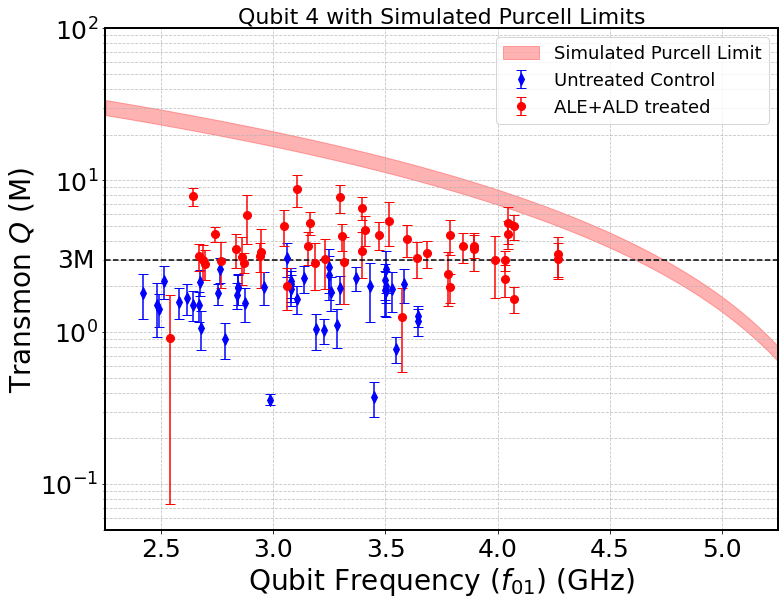}}%
    \caption{Frequency dependence of the estimated $Q$ for each qubit design, for various untreated and ALE+ALD treated devices, along with their calculated Purcell curves based on experimentally measured upper and lower limits of $\kappa_c$.}
    \label{fig:Q vs Qubit design with Purcell curve}
\end{figure}

After the ALE+ALD treatment, we consistently observed a positive frequency shift across all transmon designs. As a consequence of this frequency shift, the readout resonator Purcell loss was enhanced at the maximum $f_{01}$ transmon transition frequency. The increase in $f_{01}$ contributed to the Purcell effect, which may have been a key factor behind the observed tapering of the transmon performance at higher frequencies. To investigate this behavior further, we calculated the Purcell curves for all four transmon designs using the experimentally determined resonator coupling rate ($\kappa_c$) limits. The Purcell curves (as shown in Figure \ref{fig:Q vs Qubit design with Purcell curve}) were calculated using the following formula:
The decay time \( T_P \) is given by:

\begin{equation}
T_P = \frac{2\pi}{\kappa_c} \cdot \left( \frac{g f_{r} - f_{01}}{f_{r}} \right)^2
\end{equation}
where,

\( T_P \) is the Purcell-limited decay time of the qubit — the rate at which the qubit loses energy due to coupling to the readout resonator,

\( \kappa_c \) is the resonator coupling rate (external linewidth), typically defined as \( \kappa_c = \omega_r / Q_c \), where \( Q_c \) is the coupling quality factor, and  \(\omega_r\) is the angular resonance frequency,

\( g \) is the qubit-resonator coupling strength, describing the strength of interaction between the two systems,

\( f_r \) is the resonator frequency,

\( f_{01} \) is the qubit transition frequency, corresponding to the energy difference between the ground and first excited state.

The coupling \( g \) is defined as:
\begin{equation}
g = \frac{2\pi f_{r} f_{01}}{C_{qu} C_{rr}} \cdot \frac{C_{qu} - C_{rr}}{C_{qu-rr}}
\end{equation}
where,

\( C_{qu} \) is the total qubit capacitance,

\( C_{rr} \) is the total resonator capacitance,

\( C_{qu-rr} \) is the coupling capacitance between the qubit and resonator,

\( 2\pi f_r f_{01} \) arises from quantization of the circuit Hamiltonian and reflects the energy exchange rate between the systems.

The Purcell-limited quality factor \( Q_P \) is given by:
\begin{equation}
Q_P = \frac{2\pi f_{01}}{T_P}
\end{equation}

As suggested by the plots in Figure \ref{fig:Q vs Qubit design with Purcell curve}, the readout Purcell loss appears to contribute to loss for all four transmon designs near their maximum $f_{01}$ frequency. In future studies, we plan to investigate fabrication methods to minimize treatment-induced shifts of the transmon frequency toward the readout frequency, thereby reducing the Purcell loss in our devices.


\section{Frequency shift after ALE+ALD treatment for individual transmons}

\begin{figure} [h]
    \centering
    \includegraphics[width=0.45\textwidth]{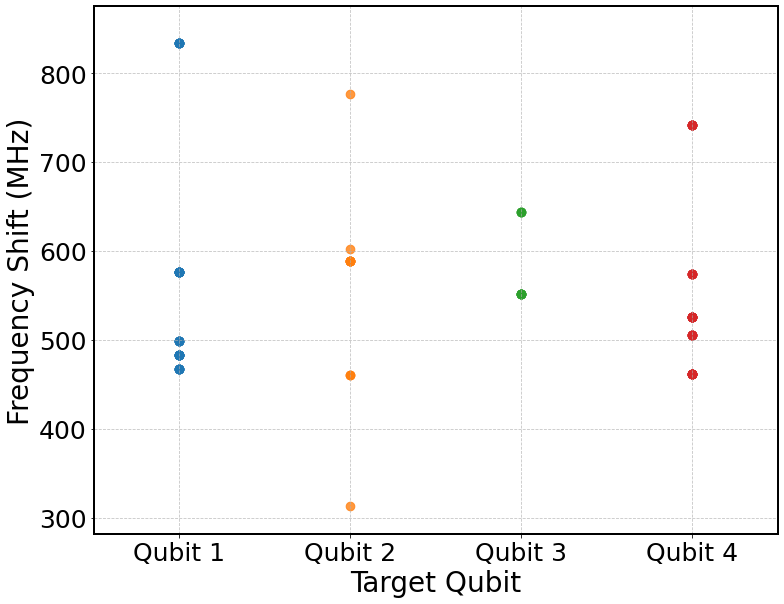}
    \caption{Measured upward frequency shift after ALE+ALD treatment on individual transmon designs, showing a consistent increase of 500-800 MHz, irrespective of the untreated transmon frequency.}
    \label{fig:freq shift per qubit}
\end{figure}

As shown in Figure \ref{fig:freq shift per qubit}, an upward frequency shift of +500-800 MHz is consistently observed post ALE+ALD across all studied transmon designs, independent of their untreated starting frequencies or source wafer. While we attribute this positive frequency shift to the high processing temperatures of ALE and ALD, the exact magnitude of this shift may be influenced by additional factors, including fabrication process variations and differences in residual polymeric contamination on the surface prior to treatment.


\section{Transmon $Q$ dependence as a function of frequency shift after ALE+ALD treatment}


\begin{figure} [htbp]
    \centering
    \includegraphics[width=0.45\textwidth]{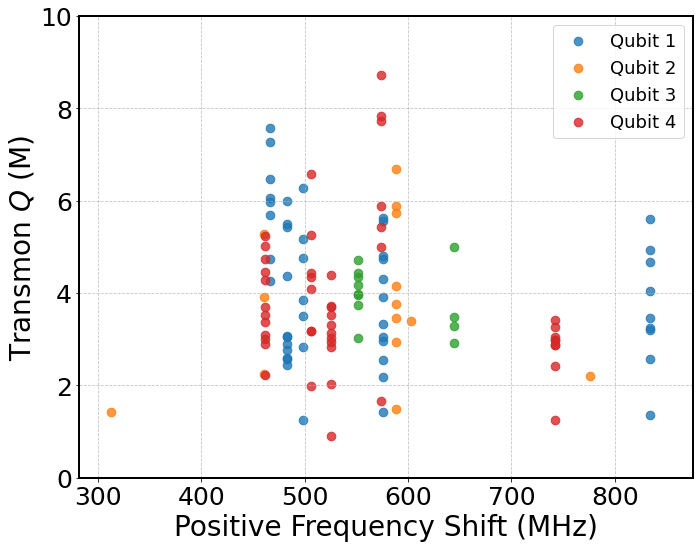}
    \caption{Measured dependence of transmon $Q$ on the absolute frequency shift for various ALE+ALD treated devices.}
    \label{fig:Q vs frequency shift}
\end{figure}

We observe a noticeable correlation between frequency shift and transmon quality factor $Q$, where the median transmon $Q$ decreases with increasing frequency shift, as shown in Figure \ref{fig:Q vs frequency shift}. This behavior is attributed to enhanced readout Purcell loss as the transmon $f_{01}$ approaches the readout frequency, as discussed in Section II of the SI. Specifically, the median transmon $Q$ decreases from approximately 4–6x10$^{6}$ for frequency shifts of 450 MHz to about 2–3x10$^{6}$ for shifts above 700 MHz.


\section{XPS peak fitting}
Surface analysis was conducted on four samples using x-ray photoelectron spectroscopy (XPS): as-deposited blanket Al film, untreated device, ALE+ALD treated device, and annealed-only device. Experimental spectra for the Al $2p$, C $1s$, and O $1s$ core levels were individually curve-fitted after subtracting a Shirley background using CasaXPS.

\begin{figure}[htbp]
    \centering
    \subfloat[]{\includegraphics[width=0.45\textwidth]{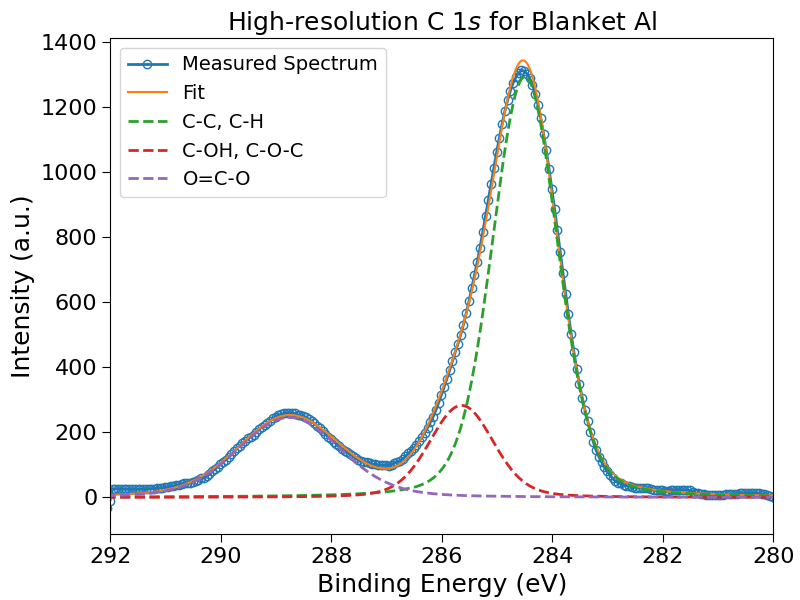}\label{fig:sub10}}\quad%
    \subfloat[]{\includegraphics[width=0.45\textwidth]{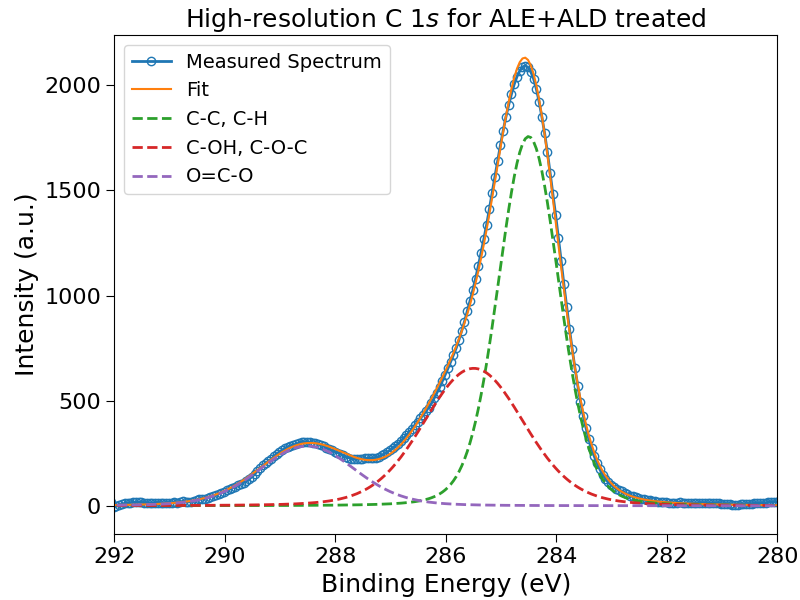}\label{fig:sub11}}\quad%
    \subfloat[]{\includegraphics[width=0.45\textwidth]{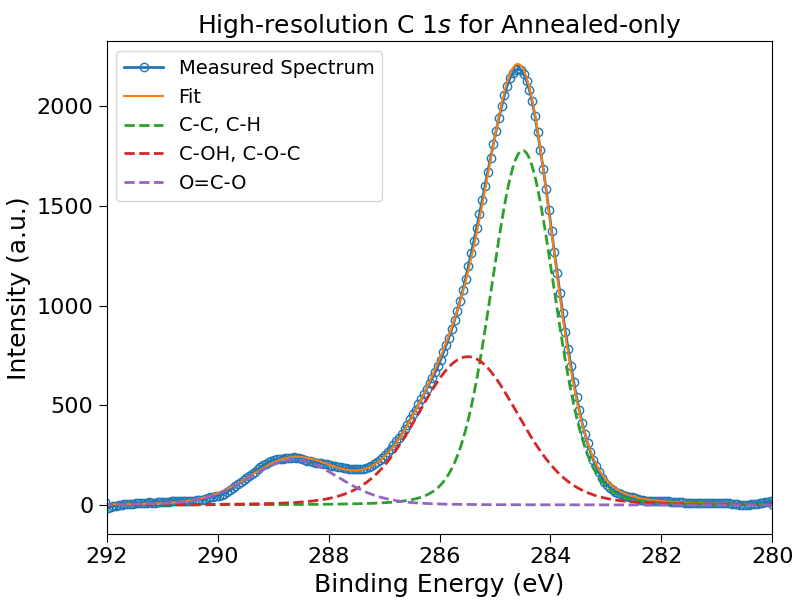}\label{fig:sub12}}\quad%
    \subfloat[]{\includegraphics[width=0.45\textwidth]{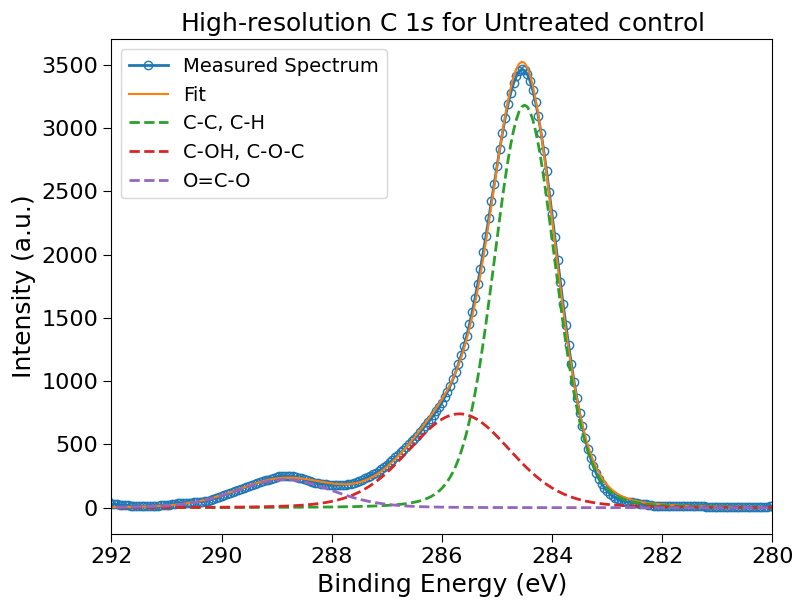}}%
     \caption{C \texorpdfstring{$1s$}{1s} core-level XPS spectra of (a) as-deposited blanket Al film, (b) ALE+ALD treated device, (c) annealed-only device, and (d) untreated device.}
    \label{fig:C 1s}
\end{figure}

\begin{figure}[H]
    \centering
    \subfloat[] {\includegraphics[width=0.45\textwidth]{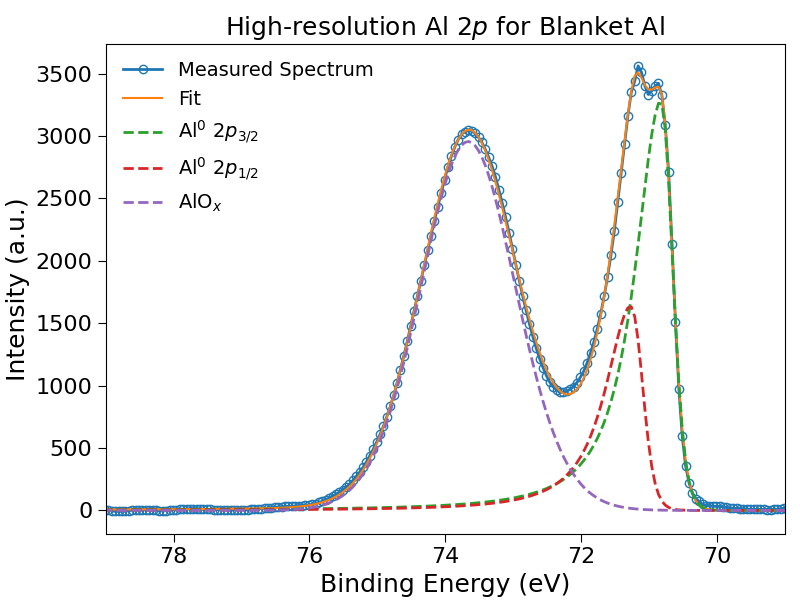}\label{fig:sub13}}\quad%
    \subfloat[] {\includegraphics[width=0.45\textwidth]{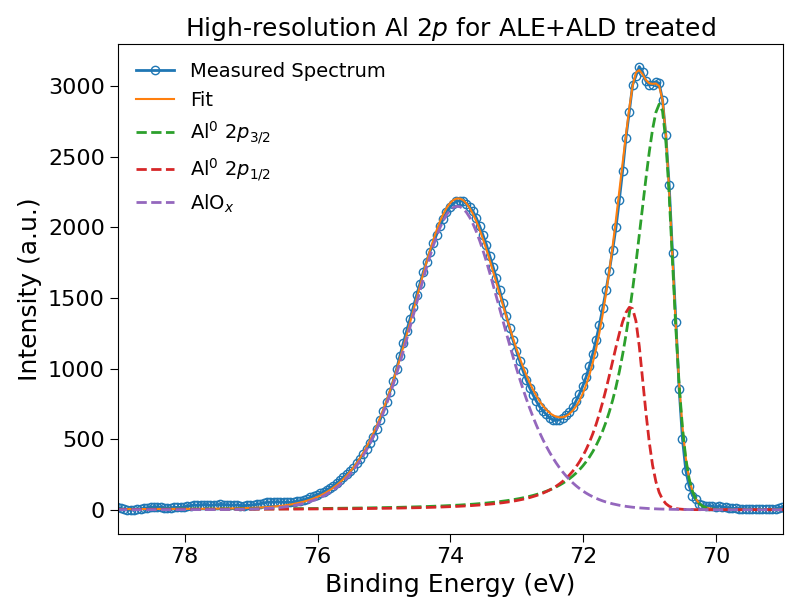}\label{fig:sub14}}\quad%
    \subfloat[] {\includegraphics[width=0.45\textwidth]{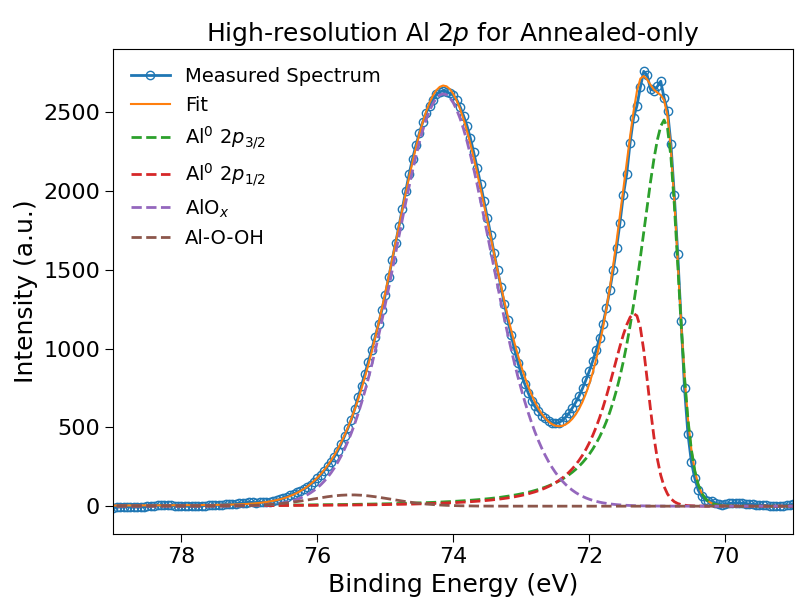}\label{fig:sub15}}\quad%
    \subfloat[]{\includegraphics[width=0.45\textwidth]{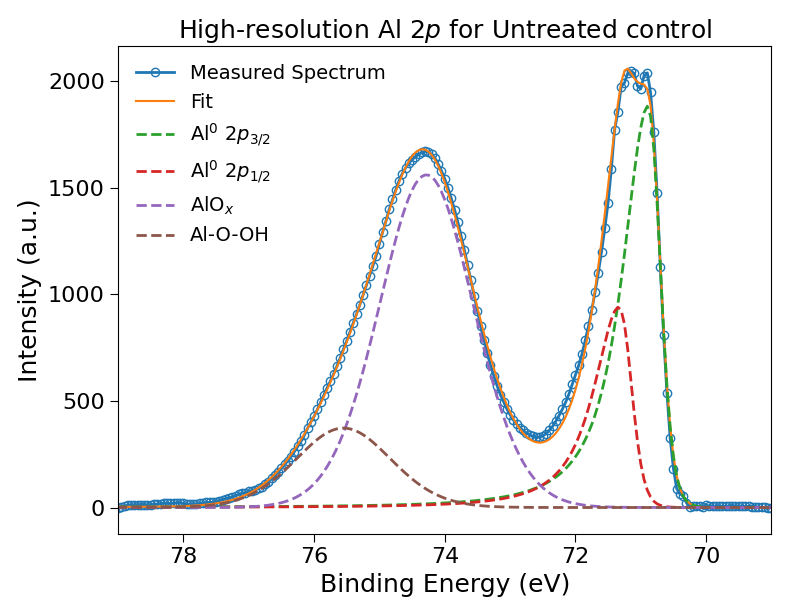}}%
    \caption{Al \texorpdfstring{$2p$}{2p} core-level XPS spectra of (a) as-deposited blanket Al film, (b) ALE+ALD treated device, (c) annealed-only device, and (d) untreated device.}
    \label{fig:Al 2p}
\end{figure}

\begin{figure}[H]
    \centering
    \subfloat[] {\includegraphics[width=0.45\textwidth]{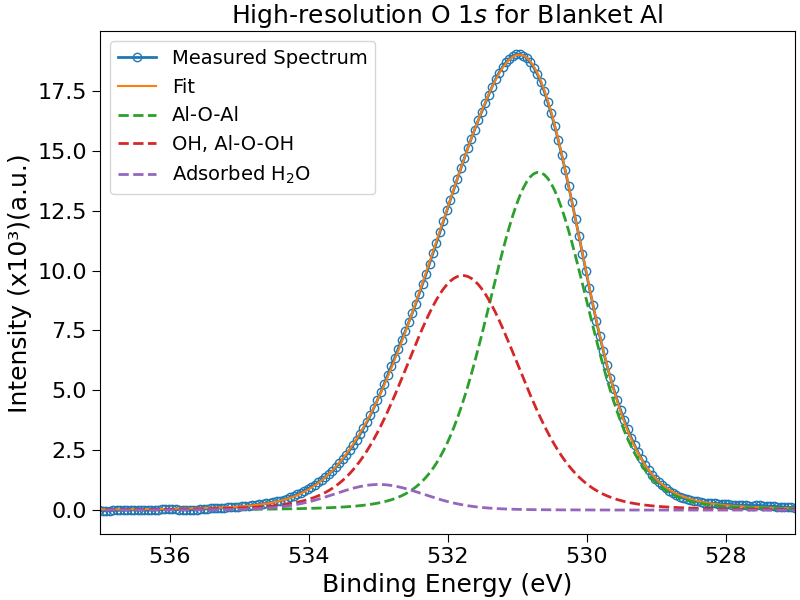}\label{fig:sub16}}\quad%
    \subfloat[] {\includegraphics[width=0.45\textwidth]{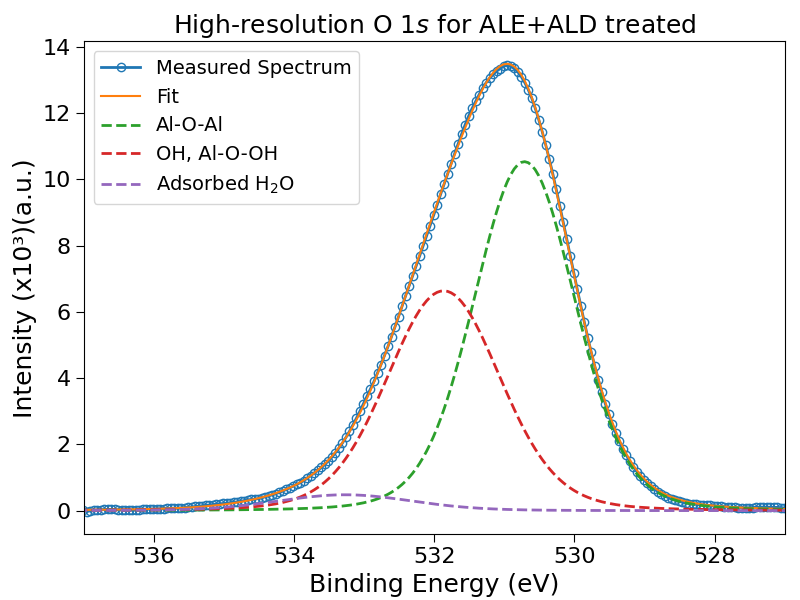}\label{fig:sub17}}\quad%
    \subfloat[] {\includegraphics[width=0.45\textwidth]{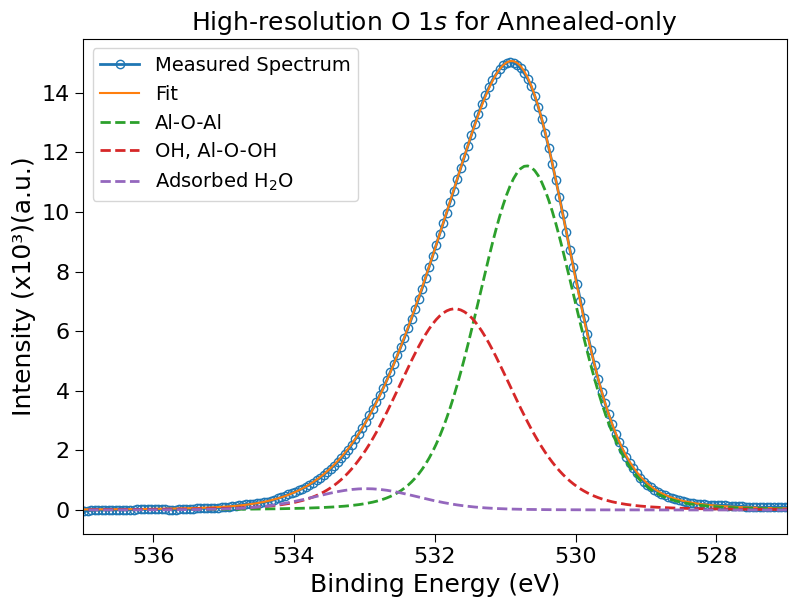}\label{fig:sub18}}\quad%
    \subfloat[]{\includegraphics[width=0.45\textwidth]{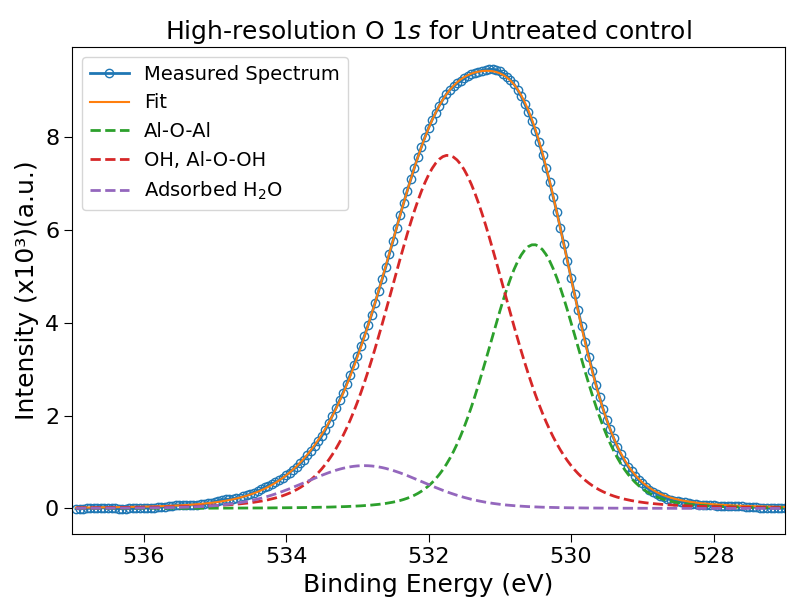}}%
    \caption{O \texorpdfstring{$1s$}{1s} core-level XPS spectra of (a) as-deposited blanket Al film, (b) ALE+ALD treated device, (c) annealed-only device, and (d) untreated device.}
    \label{fig:O 1s}
\end{figure}

A symmetric Voigt function was used for all the three components in the curve-fitting of the C $1s$ spectrum (Figure \ref{fig:C 1s}), which reveals peaks at 284.5 eV corresponding to C-C and C-H bonds from adventitious carbon, $\sim$ 285.5 eV for C-OH and C-O-C species; and $\sim$ 288.5 eV for O=C-O functional groups associated with surface oxidation or contamination. The Al $2p$ core-level spectra (Figure \ref{fig:Al 2p}) were analyzed with a symmetric Voigt function for oxides and a Lorentzian asymmetric function for metallic Al. The metallic doublet positions were constrained within 0.44 eV with a 2:1 area ratio for Al $2p_{3/2}$ and Al $2p_{1/2}$. A single oxide component was identified in the as-deposited Al film and ALE+ALD treated device, while an additional peak around 75.6 eV (Al-O-OH) appeared due to organic or polymeric contamination in the annealed-only and untreated devices. A shift of the Al oxide component for the annealed-only and untreated devices towards higher binding energies suggests the formation of oxygen-rich AlO$_{x}$. The deconvolution of the O $1s$ core-level spectra (Figure \ref{fig:O 1s}) using symmetric Voigt functions reveals the presence of three components: $\sim$ 530.7 eV for Al-O-Al bonding, $\sim$ 531.7 eV for -OH or Al-O-OH species, and $\sim$ 533 eV for adsorbed H$_{2}$O. 

\section{Estimation of \texorpdfstring{Al$_2$O$_3$}{Al2O3} thickness from XPS analysis}

The thickness of the Al$_2$O$_3$ layer was estimated based on the intensity ratio of the Al $2p$ peaks respectively corresponding to the oxide and the underlying metal in the experimental spectra. This method is based on the attenuation of photoelectrons emitted from the metal substrate as they pass through the oxide layer. The thickness $d$ of the Al$_2$O$_3$ layer can be inferred using the following equation based on the Beer-Lambert Law:

\[
d_{xps} (\text{\AA}) = \lambda_0 \sin \theta \cdot \ln \left( \frac{N_m \lambda_m}{N_0 \lambda_0} \cdot \frac{I_0}{I_m} + 1 \right)
\]

where, 

d = thickness of the Al$_2$O$_3$ layer

$\lambda$$_0$ = inelastic mean free path (IMFP) of Al $2p$ electrons in Al$_2$O$_3$ 

$\lambda$$_ m$ = IMFP of Al $2p$ electrons in metallic Al 

$\theta$ = take off angle 

I$_m$ = XPS intensity of metallic Al $2p$ peak

I$_0$ = XPS intensity of oxidized Al $2p$ peak

N$_m$ = Atomic density of metallic Al

N$_0$ = Atomic density of metallic Al$_2$O$_3$

The XPS intensity was determined from the total area under the curve of the Al $2p$ core level for the metallic and oxide components. Using the above equation with a take-off angle of 90 degrees, and $\lambda$$_0$ and $\lambda$$_m$ values of 28 $\AA$ and 26 $\AA$ for Al oxide and Al metal, respectively, along with an N$_m$:N$_0$ ratio of 1.5, the thickness was estimated for different samples and several locations on each chip as shown in Figure 3(b) of the main text.


\section{TEM-EELS analysis}
\begin{figure*}[htbp]
    \centering
    \subfloat[]{\includegraphics[width=0.23\textwidth]{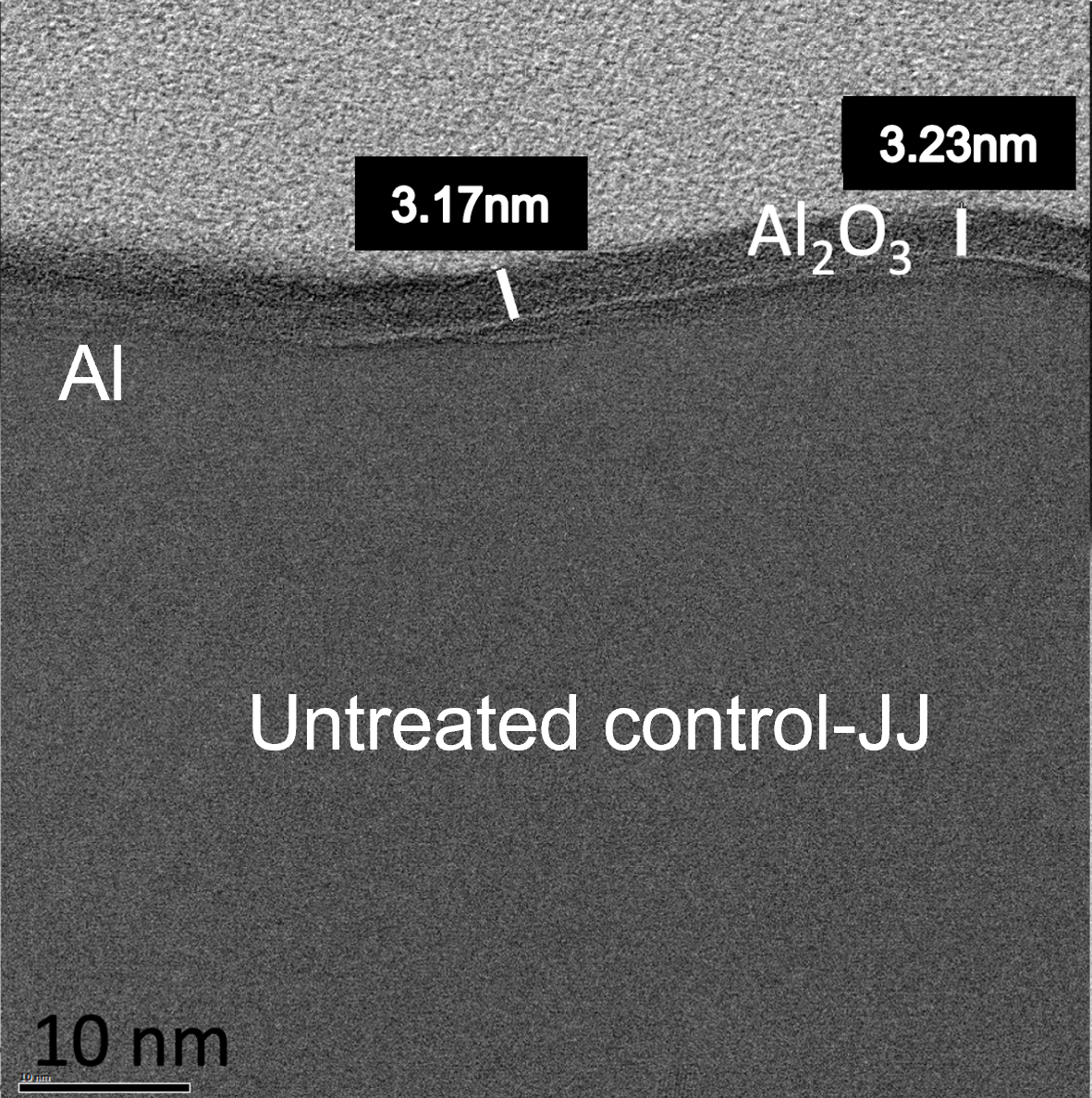}\label{fig:sub19}}\quad%
    \subfloat[]{\includegraphics[width=0.23\textwidth]{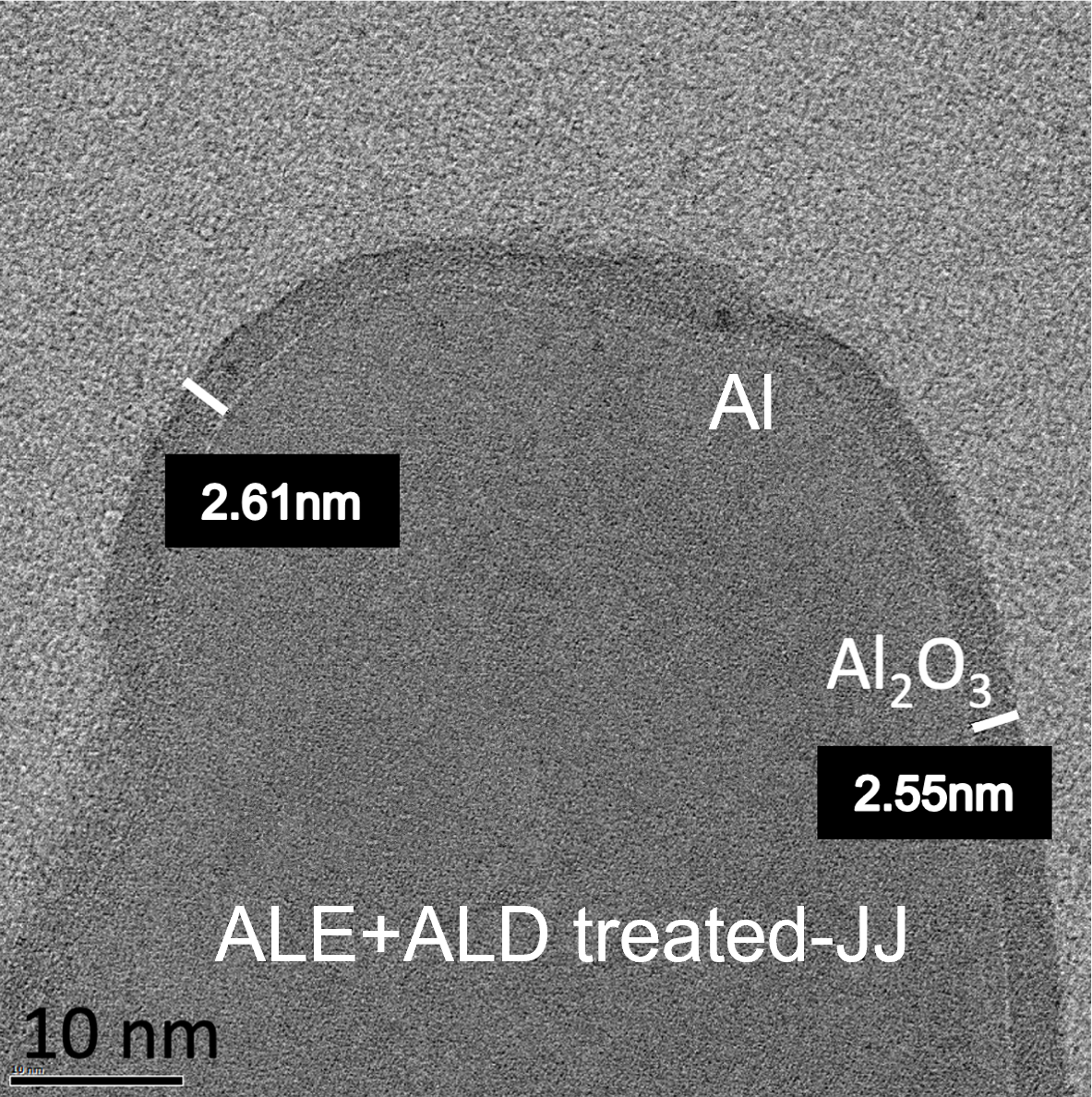}\label{fig:sub20}}\quad%
    \subfloat[]{\includegraphics[width=0.23\textwidth]{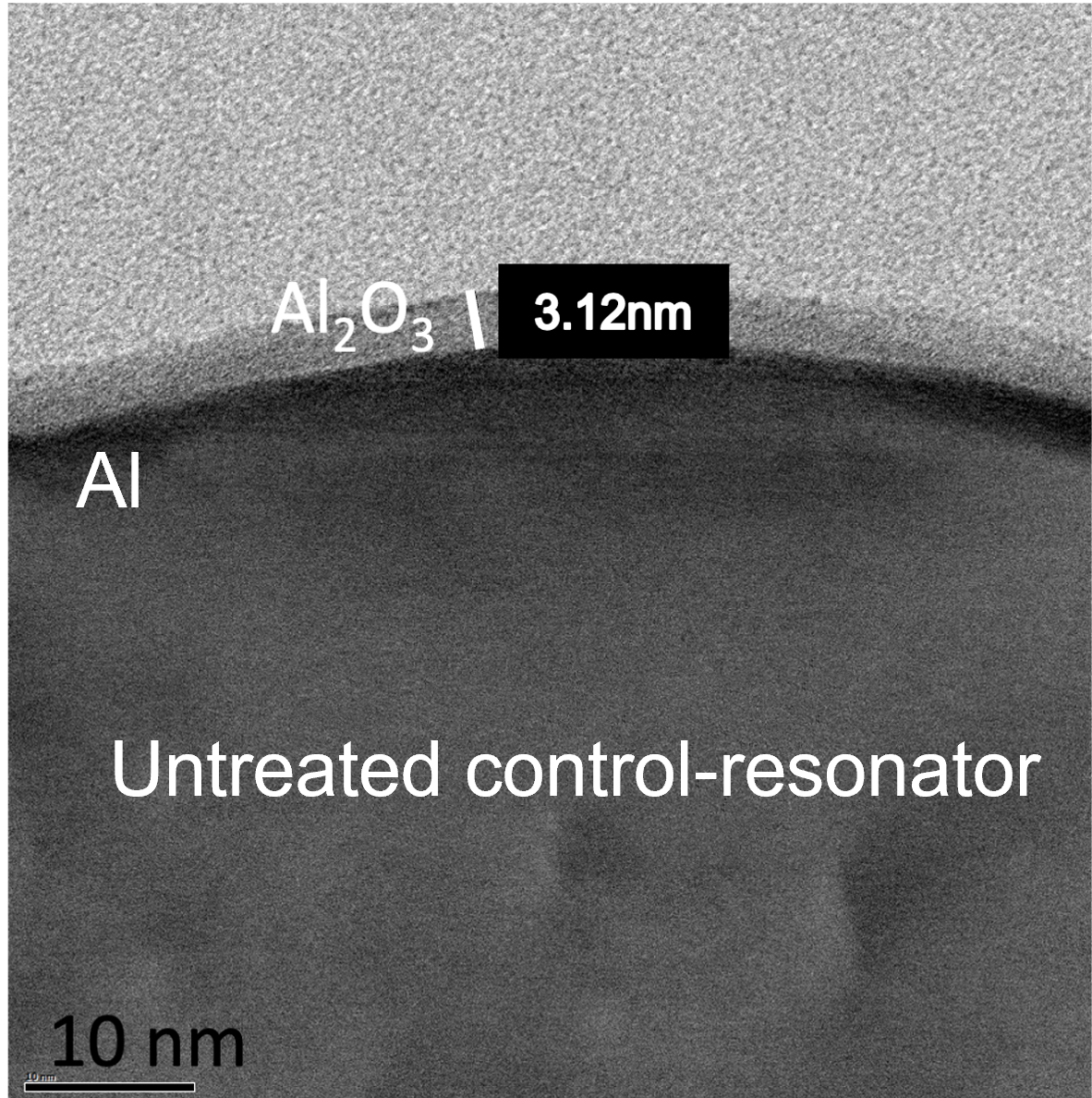}\label{fig:sub21}}\quad%
    \subfloat[]{\includegraphics[width=0.23\textwidth]{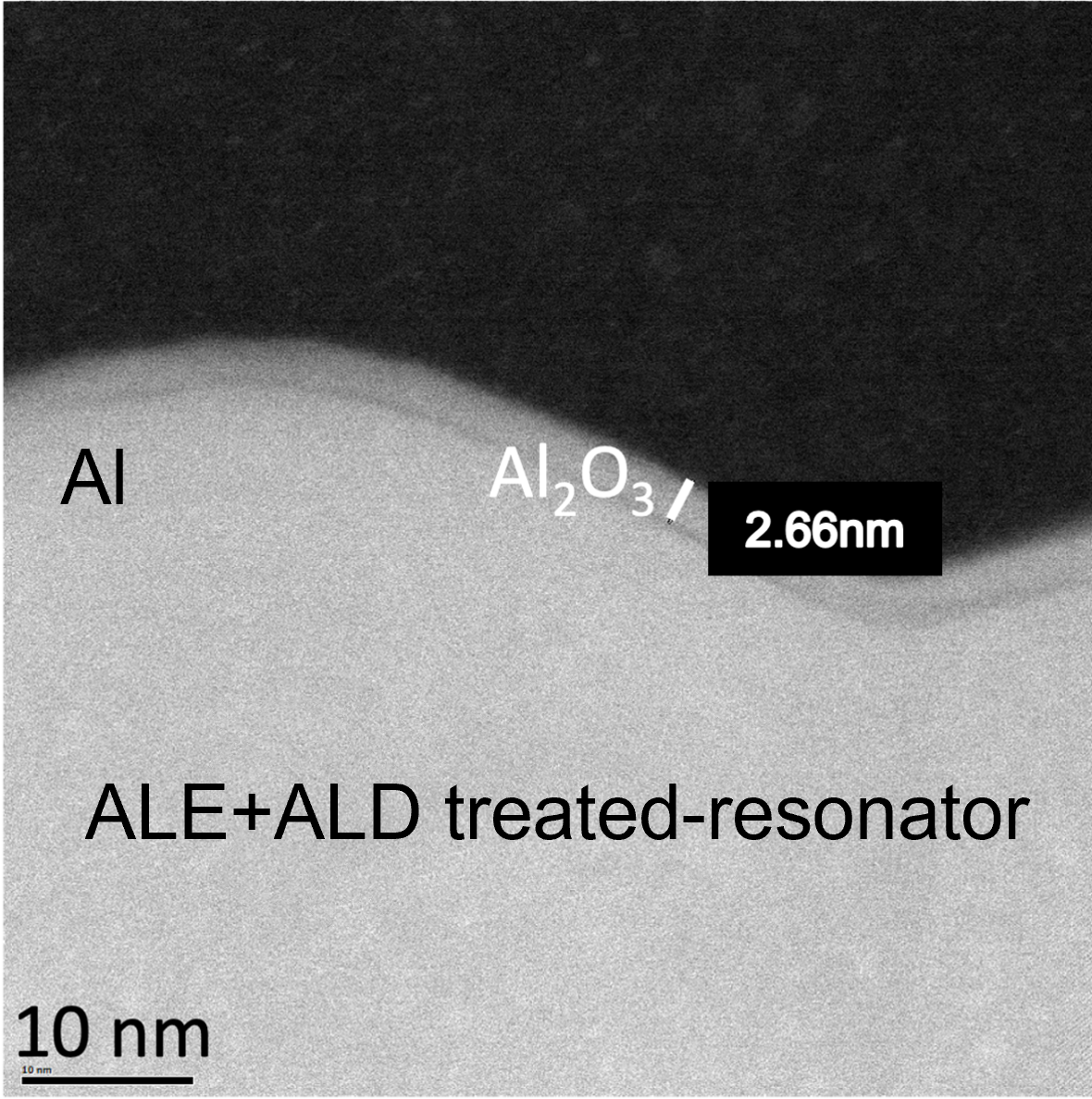}}\quad%
    \subfloat[]{\includegraphics[width=0.23\textwidth]{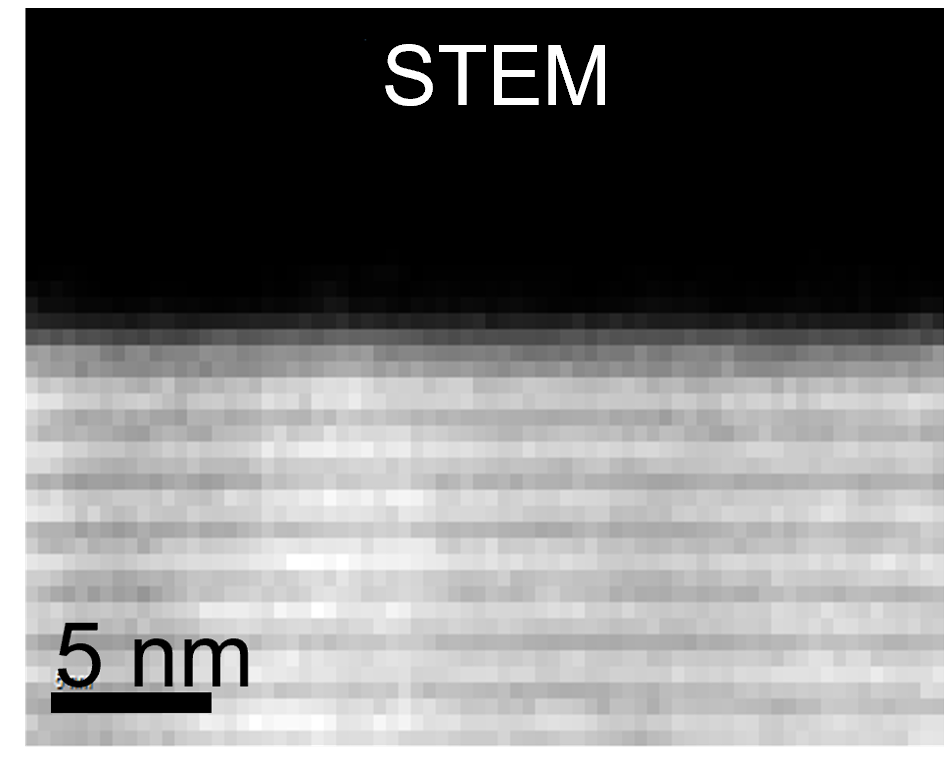}}%
    \subfloat[]{\includegraphics[width=0.23\textwidth]{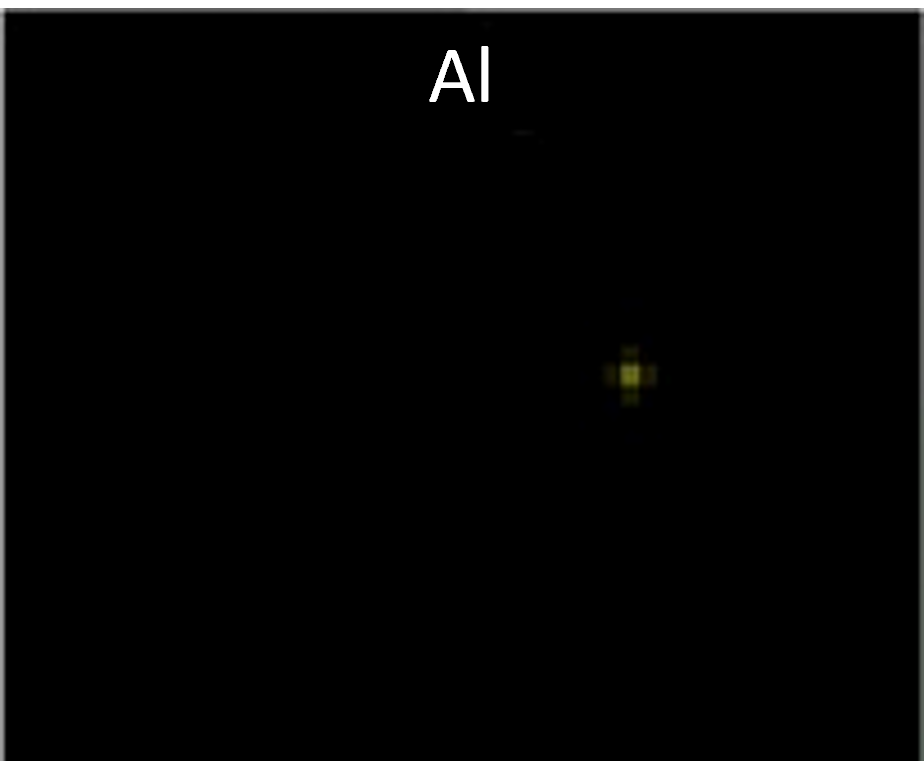}}%
    \subfloat[]{\includegraphics[width=0.23\textwidth]{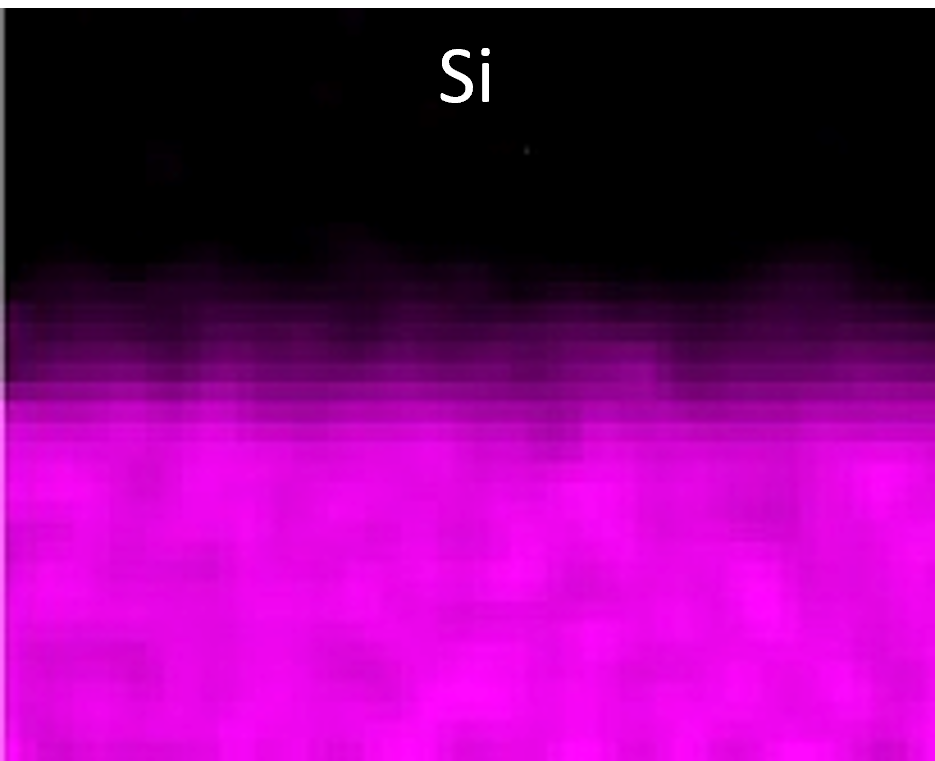}}%
    \subfloat[]{\includegraphics[width=0.23\textwidth]{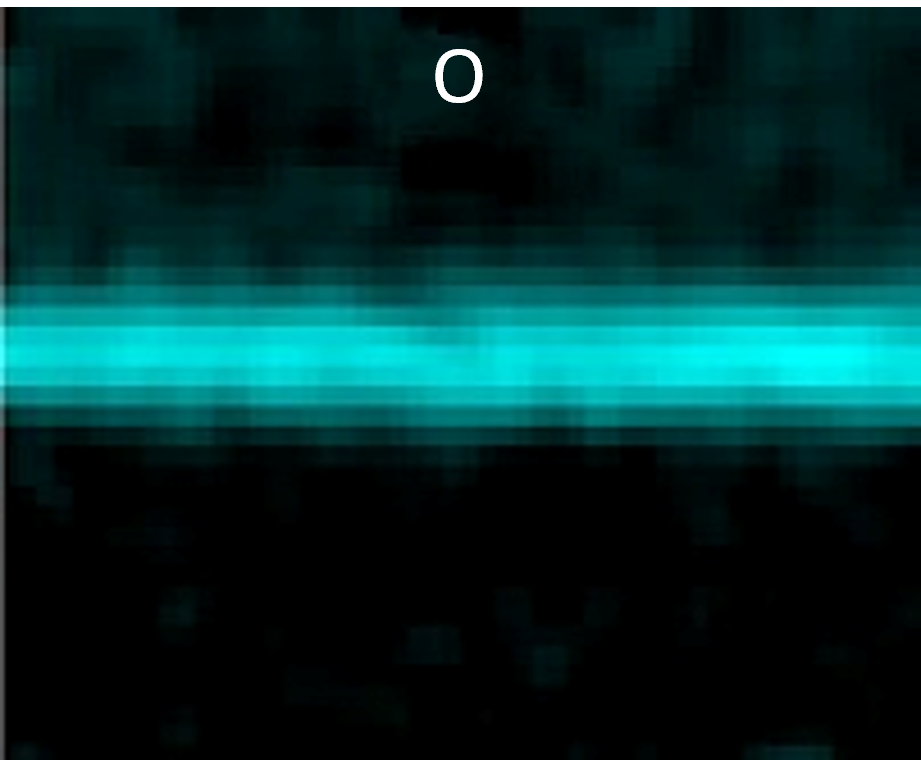}}%
    \caption{(a-d) High-resolution scanning-TEM images of the vicinity of the
surface of the the JJ top electrode and read-out resonator
trace on untreated and treated devices. (e) S-TEM image of the Si trench region in the ALE+ALD-treated resonator, alongwith TEM-EELS elemental maps for (f) Al, (g) Si and (h) O, showing the absence of detectable Al$_2$O$_3$ on the exposed Si surface.}
    \label{fig:TEM}
\end{figure*}

We performed imaging of the JJ and resonator areas to study the surface oxide layers using transmission electron microscopy and electron energy loss spectroscopy (TEM-EELS), as shown in Figure \ref{fig:TEM}(a-d). The thicknesses of the surface Al oxide layers were found to be consistent with XPS estimations: $2.5 \pm 0.2$ nm for the ALE+ALD treated chip and in the $3.1 \pm 0.2$ nm for the untreated chip. This confirms a measurable reduction in the oxide layer thickness of the treated chip, and underscores the uniformity of the treatment's effect on the surface oxide layer. The oxide thickness was estimated by averaging measurements from 25 locations across all the images shown in Figure \ref{fig:TEM} using the software ImageJ.
For each chip, the O/Al ratio was also estimated based on an EELS map collected across the surface oxide layer. This ratio was estimated to be approximately 1.5 for both chips, suggesting that the differences in oxidation state identified via XPS are either too small to be detected just by TEM-EELS spectral mapping alone or have been further minimized by the lamella preparation process used for TEM.
Additionally, EELS maps (Figure \ref{fig:TEM}(f-h)) measured on exposed Si areas next to the resonators indicate that the native silicon oxide is not capped by Al$_2$O$_3$, which is consistent with the PiFM data.

\section{IR peak assignment}
Table \ref{tab:IR_peak_assignments} summarizes the peak assignments for all peaks identified in the Al and Si regions of the resonators and junctions on both untreated and treated surfaces.

\begin{table*}[h]
    \centering
    \renewcommand{\arraystretch}{1.3} 
    \resizebox{\textwidth}{!}{ 
        \begin{tabular}{|c|c|c|c|c|c|}
            \hline
             \multirow{3}{*}{\centering\textbf{Peak Assignment}} & \multicolumn{5}{c|}{\textbf{Wavenumbers (cm$^{-1}$)}} \\
            \cline{2-6}
            & \multicolumn{2}{c|}{\textbf{Untreated}} & \multicolumn{3}{c|}{\textbf{ALE+ALD Treated}} \\
            \cline{2-6}
            & \textbf{Resonator} & \textbf{Junction} & \textbf{Resonator-Al trace} & \textbf{Resonator-Si trench} & \textbf{Junction} \\
            \hline
            Al-O stretching vibration & - & - & 800-1000 & - & 800-1000 \\
            \hline
            C-O stretching vibration from PMMA & - & - & 1020 & 1020 & - \\
            \hline
            C-O-C stretching vibration from PMMA & 1150 & 1150 & - & - & - \\
            \hline
            C-H bending vibration from Al-CH$_3$ & - & - & 1170 & 1170 & 1170 \\
            \hline
            C-O-C stretching vibration from PMMA & 1197 & 1197 & 1197 & 1197 & - \\
            \hline
            C-O-C stretching vibration from PMMA & 1240 & 1240 & - & - & - \\
            \hline
            C-O-C stretching vibration from PMMA & 1265 & 1265 & 1265 & 1265 & 1265 \\
            \hline
            CH$_x$ bending deformation & 1378 & 1378 & 1378 & 1378 & 1378 \\
            \hline
            CH$_x$ bending deformation & 1445 & 1445 & 1445 & 1445 & 1445 \\
            \hline
            CH$_x$ bending deformation & 1466 & 1466 & 1466 & 1466 & 1466 \\
            \hline
            C=O stretching vibration from PMMA & 1740 & 1740 & 1740 & 1740 & - \\
            \hline
        \end{tabular}
    }
    \caption{Infrared spectra peak assignments for untreated and ALE+ALD treated devices.}
    \label{tab:IR_peak_assignments}
\end{table*}


\section{PiFM mapping of resonator area }
\begin{figure}[h]
    \centering
    \subfloat[]{\includegraphics[width=0.5\textwidth]{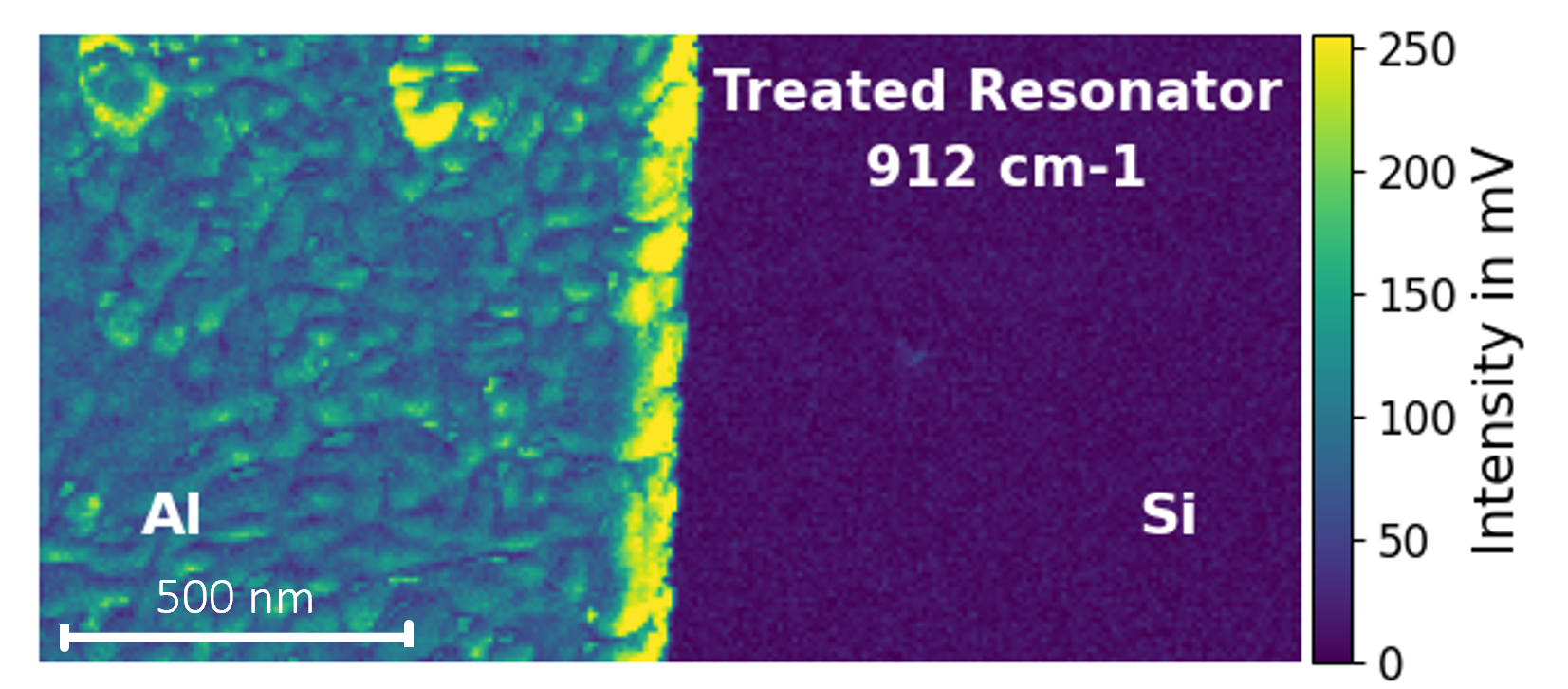}}\label{fig:sub22}%
    \subfloat[]{\includegraphics[width=0.5\textwidth]{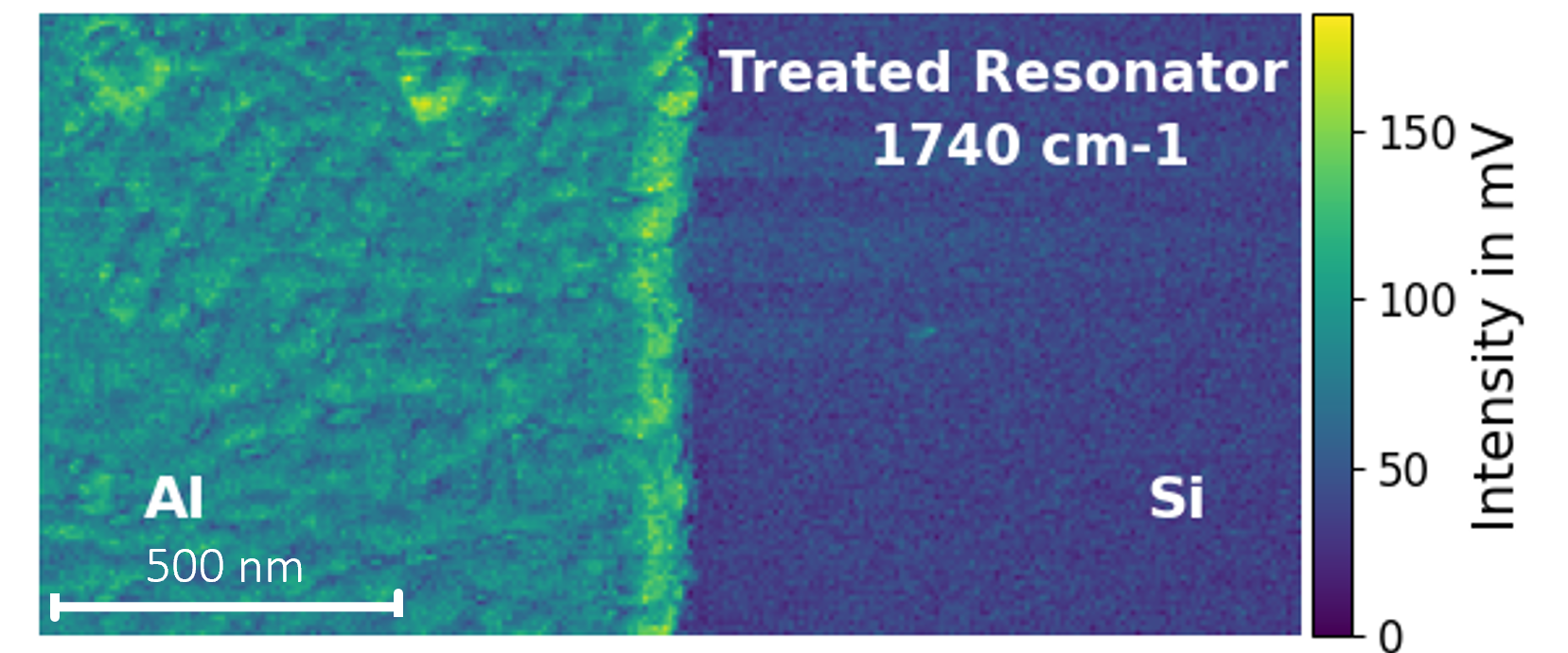}}\label{fig:sub23}%
    \caption{PiFM maps at the resontor area shown with a color scale highlighting the 912 cm$^{-1}$ and 1740 cm$^{-1}$ wavenumbers, which respectively correspond to Al-O bonds and C=O stretching vibration in PMMA}
    \label{fig:Raw PiFM maps at resonator}
\end{figure}

In this section, we present raw PiFM maps (Figure \ref{fig:Raw PiFM maps at resonator}) of the resonator area for ALE+ALD-treated devices, acquired at 912 cm$^{-1}$ and 1740 cm$^{-1}$, corresponding to the Al–O stretching vibration and the C=O stretching vibration in PMMA, respectively. 
